\documentclass[journal,10pt,twocolumn,twoside]{IEEEtran}
\usepackage{multicol}   
\usepackage{multirow}
\usepackage[pdftex]{graphicx}
\usepackage{algorithm,algpseudocode}
\usepackage[small]{caption}
\usepackage{float}
\usepackage{amssymb,amsmath}
\usepackage{graphicx}
\usepackage{subfigure}
\usepackage{xspace}
\usepackage{amsthm}
\usepackage{fancyhdr}

\usepackage{caption,setspace}
\usepackage{amssymb,amsmath}
\usepackage{bm}
\usepackage{textcomp} 
\usepackage{epstopdf}
\usepackage{bbding}
\usepackage{cite} 
\usepackage{color}
\usepackage{pifont}
\usepackage{lipsum}
\usepackage{amstext}
\usepackage{orcidlink}

\usepackage{cuted}

\graphicspath{{fig/}}

\begin{document}
\title{\huge Channel Uncertainty-Aware Robust Beamforming for RIS-Assisted RSMA Communication With Movable Antennas}
\author{Muhammad Asif \,\orcidlink{0000-0002-9699-1675}, Asim Ihsan \,\orcidlink{0000-0001-7491-7178}, Zhongliang Wang\,\orcidlink{0000-0003-4525-5798}, Irfan Muhammad \,\orcidlink{0000-0001-5703-7988}, Tong Shen \,\orcidlink{0000-0002-7650-3275}, Manzoor Ahmed \,\orcidlink{0000-0002-0459-9845}, Ertugrul Basar \,\orcidlink{0000-0001-5566-2392}, \IEEEmembership{Fellow, IEEE}, and Symeon Chatzinotas \,\orcidlink{0000-0001-5122-0001}, \IEEEmembership{Fellow, IEEE}

\thanks{This work was supported in part by the Natural Science Research Project of Anhui Educational Committee (2023AH040232, KJ2016A884-2025).}  	

\thanks{Muhammad Asif and Tong Shen are with the School of Electrical and Information Engineering, Tongling University, Anhui, Tongling, China (e-mail: masif@tlu.edu.cn; 2024118@tlu.edu.cn).
	
  Asim Ihsan is with the Interdisciplinary Research Center for Communication Systems and Sensing, King Fahd University of Petroleum \& Minerals (KFUPM ) Dhahran, Saudi Arabia (e-mail: asim.ihsan@kfupm.edu.sa).
	
  Zhongliang Wang with the School of Electrical and Information Engineering, Tongling University, Anhui, Tongling, China, and also with the Engineering Technology Research Center of Optoelectronic Technology Appliance, Anhui Province, Tongling, China (e-mail:asdwzl@tlu.edu.cn).
  
  Manzoor Ahmed is with the School of Computer and Information Science and the Institute for AI Industrial Technology Research, Hubei Engineering University, Xiaogan 432000, China (e-mail: manzoor.achakzai@gmail.com).
	
  Irfan Muhammad is with the Centre for Wireless Communications, University of Oulu, 90570 Oulu, Finland (e-mail: irfan.muhammad@oulu.fi).
  
  Ertugrul Basar is with Tampere Wireless Research Center, Department of Electrical Engineering, Tampere University, 33720 Tampere, Finland, and also with the Department of Electrical and Electronics Engineering, Koc University, 34450 Istanbul, Turkiye (e-mail: ertugrul.basar@tuni.fi).

  Symeon Chatzinotas is with the Interdisciplinary Centre for Security, Reliability and Trust (SnT), University of Luxembourg, 1855 Luxembourg City, Luxembourg (e-mail: symeon.chatzinotas@uni.lu).

}

\vspace{-0.7cm}}%

\markboth{}
{ \MakeLowercase{\textit{}}} 
\maketitle

\begin{abstract}
This work investigates a robust resource allocation framework for a downlink multi-user communication system integrating movable antennas (MAs) and reconfigurable intelligent surface (RIS) under the rate-splitting multiple access (RSMA) transmission protocol. Unlike conventional fixed-position antenna architectures, the considered MAs-enabled system introduces spatially adaptive channel variations in which antenna positions directly influence the effective channel responses. Under imperfect channel state information (CSI), changes in the MAs positions and RIS reflection coefficients affect both the estimated effective channels and the corresponding uncertainty radii under the adopted uncertainty model. This dependence leads to a highly coupled robust optimization problem. To address this challenge, we formulate a system sum-rate maximization problem by jointly optimizing the transmit precoding vectors, RIS reflection matrix, common-rate allocation, and MAs positions, subject to quality-of-service (QoS), power-budget, common-rate decoding, and minimum antenna-separation constraints. The resulting non-convex problem is efficiently handled through an iterative robust optimization framework, where the original problem is successively decomposed into active beamforming, RIS reflection matrix, and MAs position optimization subproblems, and tractable convex surrogate functions are constructed to enable iterative optimization. Moreover, conservative bounds on the common- and private-stream rates are derived under the bounded CSI uncertainty model to enforce the rate requirements within the adopted uncertainty sets. Finally, extensive simulation results demonstrate that the proposed framework achieves significant performance gains and enhanced robustness compared with benchmark schemes, while also exhibiting fast and stable convergence behavior under practical imperfect CSI conditions.

\end{abstract}


\IEEEpeerreviewmaketitle


\section{Introduction}

\IEEEPARstart{T} {he} rapid evolution of wireless communication systems has fueled growing interest in the development of sixth-generation (6G) networks \cite{wang2023road,saad2019vision}. The future 6G communication systems are likely to offer new performance requirements such as high data rates, high reliability, and better spectral efficiency, which will require new transmission technologies to be explored \cite{matthaiou2021road,wang2022gcwcn,chen2020vision}. In this context, Multi-antenna communication technologies, especially multiple-input multiple-output (MIMO) systems, are among the most promising schemes that have been extensively analyzed due to their ability to exploit spatial multiplexing for improving both spectral and energy efficiency \cite{di2011spatial,molisch2017hybrid}. Moreover, in future wireless networks, the growing adoption of high-frequency bands with shorter signal wavelengths facilitates the practical implementation of large antenna arrays and enables massive MIMO systems to achieve substantial beamforming and spatial multiplexing gains \cite{zhu2019millimeter,larsson2014massive,ding2020joint}.

However, the traditional MIMO architectures rely on antennas deployed at fixed locations, which fundamentally limits the spatial degrees of freedom available for diversity and multiplexing gains. Recently, a new frontier technology known as movable antennas (MAs) has been introduced, providing a novel paradigm in which antennas can dynamically adjust their positions within a small region to fully exploit spatial degrees of freedom, resulting in improved communication performance \cite{zhu2023movable, zhu2023modeling}. Specifically, in MAs-assisted systems, each antenna element is connected to its corresponding radio frequency (RF) chain through a flexible cable, allowing it to move freely within a predefined spatial region, thereby enabling more favorable channel conditions and improved system performance. In addition, MAs-enables systems can effectively mitigate multiuser interference by repositioning themselves from locations with strong interference to positions where interference is weaker. By independently adjusting the locations of different MAs, the spatial adjustment of antennas enables a more effective exploitation of spatial degrees of freedom, thereby improving diversity and multiplexing performance. In this context, recent studies have demonstrated that MAs-assisted systems achieve superior communication performance compared to fixed-position antennas (FPA)-based systems. Authors of \cite{zhu2023modeling} introduced the concept of MAs and established a far-field channel model based on the field-response formulation. Using this model, they analyzed the maximum achievable channel gain of MAs-assisted systems with a single movable antenna deployed at both the transmitter and receiver nodes. Their results demonstrated that MAs systems can significantly outperform FPA counterparts, especially in rich multipath environments. Xiu et al. in \cite{xiu2026meta} investigated MAs-aided full-duplex cell-free radar--communication systems under practical synchronization impairments caused by carrier frequency offset and developed a hybrid optimization and meta-reinforcement learning framework for joint antenna positioning and beamforming. Authors of \cite{xiu2026robust} investigated MAs-aided cell-free integrated sensing and communication (ISAC) under time-synchronization errors, combining manifold optimization with meta-reinforcement learning to improve worst-case sensing accuracy. In \cite{xiao2024multiuser}, the authors investigated MAs-assisted multiuser uplink communications by proposing a base station architecture equipped with multiple movable antennas. They formulated a max–min rate optimization problem by jointly optimizing the antenna positions, receive combining, and users’ transmit powers under practical constraints. Likewise, the work in \cite{zhu2023movable123} investigated beamforming with movable antenna arrays to overcome the inherent trade-off between array gain and interference suppression in FPA systems. By jointly optimizing the antenna position vector and antenna weight vector, it is demonstrated that full array gain toward the desired direction and null steering toward multiple interference directions can be simultaneously achieved. In \cite{xiu2025movable}, authors considered imperfect CSI and time-synchronization errors in MAs-aided cooperative ISAC networks, employing constrained deep reinforcement learning for joint beamforming, BS selection, and antenna positioning to minimize power consumption. Further, Ma {et al.} in \cite{ma2023mimo} investigated a MAs-enabled point-to-point MIMO systems for capacity enhancement through antenna position optimization. By jointly optimizing the positions of MAs and transmit beamforming, the authors characterized the capacity gains achievable with MAs over FPA systems. For low-altitude airborne vehicles, the authors of \cite{xiu2025movable123} developed a block coordinate descent algorithm to jointly optimize beamforming and antenna positions for communication-capacity maximization under a sensing SNR constraint. Further, authors in \cite{zhu2024performance} investigated a MAs-enabled wide-band communication system by considering OFDM transmission over frequency-selective fading channels. A multi-tap field-response channel model is employed to characterize the joint spatial–frequency channel variations induced by antenna movement. 

Recently, rate-splitting multiple access (RSMA) has been widely investigated as a non-orthogonal physical-layer transmission paradigm, due to its superior spectral efficiency and enhanced ability to manage interference \cite{mao2022rate}. RSMA operates by splitting each user’s message into common and private components, which enables partial decoding of multiuser interference while treating the remaining interference as noise. Through proper message splitting and power control between the common and private messages, RSMA is able to dynamically support a broad spectrum of interference management strategies ranging from complete interference treatment as noise to complete interference decoding \cite{clerckx2023primer}. Owing to this natural flexibility, RSMA is able to fill the existing gap between space-division multiple access (SDMA), which is based on treating interference as noise, and non-orthogonal multiple access (NOMA), which focuses on decoding dominant interference signals. Thus, the balanced transmission system of RSMA supports enhanced spectral and energy efficiency \cite{camana2022rate}, while also providing increased robustness in multi-user communication environments characterized by adverse operating conditions. The benefits of RSMA have also motivated its application to networks with stringent delay and reliability requirements. In this context, Chen \textit{et al.}~\cite{chen2025performance} investigated resource allocation in RSMA-assisted uplink industrial Internet-of-Things networks and developed a stochastic-network-calculus framework for short-packet size maximization and transmit-power minimization under statistical QoS requirements.

Despite RSMA offers superior spectral efficiency and interference mitigation capabilities, blockage of the direct communication link can substantially impair its performance, giving rise to coverage holes and increased outage probability. To overcome this limitation, reconfigurable intelligent surface (RIS) has emerged as an effective solution that enhances capacity, reliability, and coverage by dynamically shaping the wireless propagation environment\cite{liu2021reconfigurable}. An RIS typically consists of a planar array composed of a large number of passive reflecting elements. By adaptively adjusting the reflection coefficients of these elements, the RIS can intelligently manipulate the wireless propagation environment  \cite{asif2024securing}. In particular, through optimized reflection control, the RIS is able to steer incident signals toward intended users, thereby reshaping the channel conditions and alleviating coverage blind spots\cite{ihsan2022energy, asif2025noma}. As a result, due to its largely passive operation, with power consumption mainly associated with the control circuitry, RIS offers a cost-effective, low-maintenance, and easily deployable solution.

Inspired by the aforementioned advantages, the deployment of MAs in conjunction with RSMA transmission and RIS-assisted propagation control is expected to offer a promising approach for achieving adaptive spatial control, high spectral efficiency, and reliable coverage in future 6G systems. In this regard, authors in \cite{amhaz2026enhancing} investigated a downlink coordinated multi-point RSMA system in which users are equipped with MAs to exploit spatial adaptability. A joint optimization framework was considered to maximize the system sum-rate by optimizing base-station (BS) beamforming, common stream allocation, and MAs positions under quality-of-service (QoS) constraints. Further, Li {et al.} studied a MAs-assisted RSMA framework for short-packet ultra-reliable and low-latency communication transmission. A joint beamforming and antenna position optimization problem was formulated to maximize the achievable sum rate while satisfying stringent latency and reliability requirements \cite{li2025movable}. In \cite{fang2026toward}, authors investigated a MAs-aided uplink RSMA framework for supporting mobile broadband reliable low-latency communication under short-packet transmission. A sum-rate maximization problem was formulated by jointly optimizing antenna positions, receive combining, power allocation, and SIC decoding order. Likewise, the work in \cite{wu2025movable} studied a MAs-aided RIS-assisted integrated sensing and communication (ISAC) system to enhance communication and sensing performance in coverage-limited regions. A max–min beam-pattern gain optimization problem was considered by jointly optimizing the BS covariance matrix, RIS reflection coefficients, and MAs positions under practical constraints. Ma {et al.} in \cite{ma2025movable} investigated a MAs-assisted RIS-enabled ISAC system with a focus on enhancing physical-layer security against eavesdropping. A sum-rate maximization problem was formulated by jointly optimizing BS beamforming, RIS reflection coefficients, and MA positions. The authors in \cite{sun2024sum} considered an RIS-aided multiuser communication system enhanced by MAs to improve channel capacity. A joint optimization problem was formulated to characterize system performance by optimizing the RIS reflection coefficients, MA positions, and BS beamforming, and an efficient fractional programming–based algorithm was proposed to solve the resulting non-convex problem. In \cite{xie2024movable}, covert rate maximization problem was investigated by jointly optimizing MAs movement trajectories, transmit beamforming, and RIS phase shifts. Moreover, authors in \cite{sun2025star} studied a simultaneously transmitting and reflecting RIS–aided multiuser communication system enhanced by MAs to improve sum-rate performance under statistical channel-state information (CSI). A joint optimization problem was formulated to optimize STAR-RIS transmission and reflection coefficients, BS beamforming, and MA positions.

It is worth noting that the studies in \cite{amhaz2026enhancing,li2025movable,fang2026toward} have proposed various optimization strategies for MAs-assisted RSMA networks. In addition, the works in \cite{wu2025movable,ma2025movable,sun2024sum,xie2024movable,sun2025star} have developed several optimization frameworks for MAs-enabled RIS-empowered networks. Nevertheless, robust resource allocation frameworks for RIS-empowered MAs-enabled RSMA systems under practical CSI uncertainty conditions remain largely unexplored. Hu \textit{et al.}~\cite{hu2026sum} jointly optimize the transmit precoders, common-rate allocation, RIS reflection coefficients, and MAs positions to maximize the achievable sum-rate, treating the effective channels as perfectly known. The central challenge is to design the transmit precoders, MAs positions, and RIS reflection coefficients under imperfect CSI when the spatial configuration affects both the estimated channels and their uncertainty bounds. Moving an antenna changes the phases of the propagation paths, while adjusting the RIS reflection coefficients changes how the reflected signals combine at each user. Since the same antenna positions and RIS coefficients serve all users, a change that strengthens the desired signal at one user can also increase interference at another. Under the adopted uncertainty model, these changes can also alter the absolute uncertainty radii even when the relative uncertainty level is fixed. The spatial configuration therefore affects both the signal and interference powers calculated from the estimated channels and the margins used to bound them. As a result, the spatial configuration that maximizes the sum-rate evaluated using the estimated channels may differ from the one that maximizes the conservative sum-rate under CSI uncertainty. The resulting signal and interference power bounds affect whether the design can meet each user's minimum rate requirement under channel uncertainty. They also affect the total common rate available for allocation, which is limited by the lowest conservative common-stream decoding rate across users. To account for these dependencies, we jointly optimize the transmit precoders, common-rate allocation, RIS reflection coefficients, and MAs positions while evaluating the conservative rates using the uncertainty bounds associated with each spatial configuration. Specifically, we express the uncertainty radii as linear functions of the lifted reflection matrix in the RIS update. In the MAs position update, we account for how the uncertainty radii vary with antenna position when computing the gradients and curvature bounds used to construct the rate surrogates.

 This robust design challenge motivates the present work, whose main contributions are summarized below.
 
 \begin{itemize}
    \item We formulate a worst-case sum-rate maximization problem for RIS-assisted RSMA with MAs. Under the adopted CSI uncertainty model, both the estimated effective channels and the corresponding uncertainty radii depend on the MAs positions and RIS reflection coefficients. The joint design includes the transmit precoders, common-rate allocation, RIS reflection coefficients, and MAs positions, with constraints that ensure common-stream decodability and satisfy each user's QoS requirement under worst-case CSI errors.
 	
    \item We derive explicit lower bounds on the worst-case common- and private-stream rates under spectral-norm-bounded CSI errors. These rate lower bounds are expressed in terms of the estimated effective channel matrices, the uncertainty radii associated with the MAs positions and RIS reflection coefficients, and the transmit covariance matrices. The resulting rate expressions are used to evaluate the conservative sum-rate and enforce the common-stream decoding and user minimum rate requirements.
 	
 	\item We develop an alternating optimization algorithm for transmit precoding, RIS reflection, and MAs position optimization, with the common-rate allocation updated in each stage. In the precoding update, we construct convex surrogates for the rate constraints while holding the estimated effective channels and uncertainty radii fixed. In the RIS update, we express the uncertainty radii as linear functions of the lifted RIS reflection matrix. In the MAs position update, we construct concave quadratic rate surrogates using gradients and curvature bounds that account for how both the estimated effective channels and the uncertainty radii vary with antenna position.
 	
	\item Finally, extensive simulation results demonstrate that jointly exploiting the spatial flexibility of movable antennas, the passive beamforming capability of RIS, and the interference management benefits of RSMA provides substantial gains in spectral efficiency and robustness compared with benchmark schemes, while also exhibiting fast and stable convergence behavior.
\end{itemize}

 \begin{figure}[!t]
	\centering
	\includegraphics [width=0.45\textwidth]{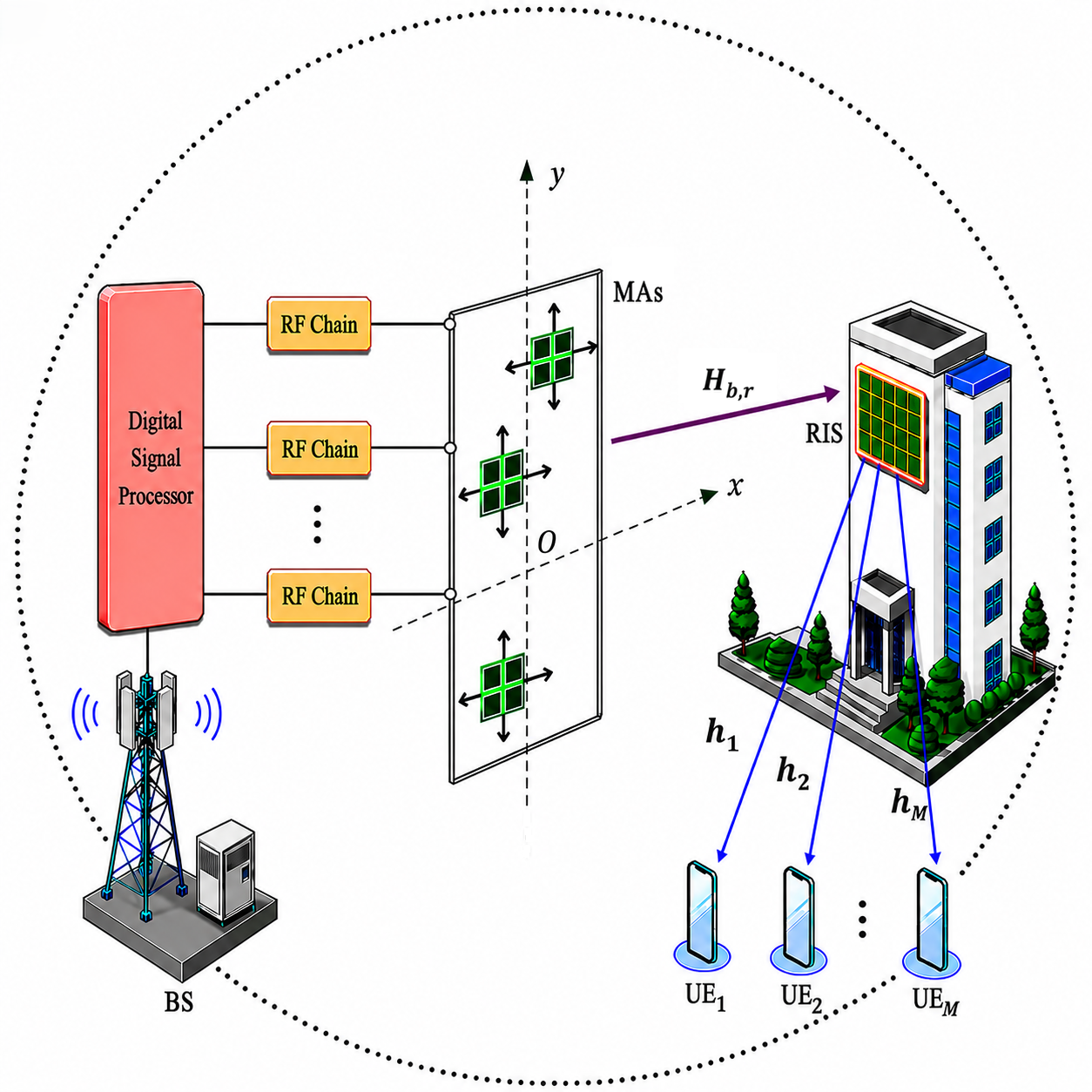}
	\caption{Illustration of system model.}
	\label{f1}
\end{figure}

\section{System Model and Optimization Problem Formulation}
As depicted in Fig.~\ref{f1}, we consider a downlink system where a BS equipped with $L$ MAs serves $M$ single-antenna RSMA users under channel estimation errors. In the presence of environmental obstructions, the direct BS--user communication links are blocked, such that user transmission is realized exclusively through the RIS equipped with $N$ reflecting elements. 

\subsection{Channel Model}
In this work, we adopt a planar far-field response channel model \cite{zhu2023modeling}, where the spatial extent of the MA transmit region is sufficiently small relative to the propagation distance between the transmitter and the receiver. Specifically, under planar far-field response model, the transmit/receive path of same channel experiences same angle-of-departure (AoD), angle-of-arrival (AoA), and amplitude of path response coefficient, whereas the phase of each path response coefficient varies across different MAs positions due to their distinct spatial locations. 

Next, let us define the positions of $L$ MAs as $\widehat{\boldsymbol{t}}=\left[\widehat{\boldsymbol{t}}_1, \widehat{\boldsymbol{t}}_2, \ldots, \widehat{\boldsymbol{t}}_L\right] \in \mathbb{R}^{2 \times L}$, where $\widehat{\boldsymbol{t}}_l=[x_l, y_l]^T\in\mathcal C_t$ denotes the position of $l$-th MA, $\forall l\in \mathcal L=\{1,2,..., L\}$, $\mathcal C_t$ denotes the transmit region of $L$ MAs at BS. Further, let $L_t$ and $L_r$ denote the numbers of transmit and receive paths, respectively. The elevation and azimuth angles at the BS and the RIS are denoted by $\phi_i^{{e}}\in[0,\pi]$ and $\phi_i^{{a}}\in[0,\pi]$, for $1\leq i\leq L_t$, and $\varphi_j^{{e}}\in[0,\pi]$ and $\varphi_j^{{a}}\in[0,\pi]$, for $1\leq j\leq L_r$, respectively. Then, the difference of the signal propagation distance for $i$-th transmit path between $\widehat{\boldsymbol{t}}_l$ and reference origin is given as
\begin{equation}
	\chi\left(\widehat{\boldsymbol{t}}_l, \phi_i^{{e}}, \phi_i^{{a}}\right)=x_l \sin \phi_i^{{e}} \cos \phi_i^{{a}} + y_l \cos \phi_i^{{e}}. \label{1}
\end{equation}
Then, for BS-RIS link, the field response vector (FRV) of MA at $\widehat{\boldsymbol{t}}_l$ is given as
\begin{equation}
	\boldsymbol{f^t}\left(\widehat{\boldsymbol{t}}_l\right)=\left[e^{j\frac{2 \pi}{\lambda} \chi\left(\widehat{\boldsymbol{t}}_l, \phi_1^e, \phi_1^a\right)}, \ldots, e^{j \frac{2 \pi}{\lambda} \chi\left(\widehat{\boldsymbol{t}}_l, \phi_{L_t}^{e}, \phi_{L_t}^{a}\right)}\right]^T\in \mathbb{C}^{L_t \times 1}, \label{2}
\end{equation} 
where $2\pi \chi(\widehat{\boldsymbol{t}}_l,\phi_i^{\mathrm{e}},\phi_i^{\mathrm{a}})/\lambda$, 
for $1\leq i\leq L_t$, denotes the phase difference of the $i$-th transmit path between the MA position $\widehat{\boldsymbol{t}}_l$ and the reference origin, and $\lambda$ represents the carrier wavelength. Accordingly, the field response matrix (FRM) for BS-RIS communication link for all $L$ MAs is given as

\begin{equation}
	\boldsymbol{F^t}(\widehat{\boldsymbol{t}}) \triangleq\left[\boldsymbol{f^t}(\widehat{\boldsymbol{t}}_1), \boldsymbol{f^t}(\widehat{\boldsymbol{t}}_2), \ldots, \boldsymbol{f^t}(\widehat{\boldsymbol{t}}_L)\right] \in \mathbb{C}^{L_t \times L}. \label{3}
\end{equation}

Similarly, the FRV of $n$-th reflecting element of RIS at position $\widehat{\boldsymbol{r}}_n=[x_n, y_n]^T$ is given by
\begin{equation}
	\boldsymbol{f^r}\left(\widehat{\boldsymbol{r}}_n\right)=\left[e^{j\frac{2 \pi}{\lambda} \chi\left(\widehat{\boldsymbol{r}}_n, \varphi_1^e, \varphi_1^a\right)}, \ldots, e^{j \frac{2 \pi}{\lambda} \chi\left(\widehat{\boldsymbol{r}}_n, \varphi_{L_r}^{e}, \varphi_{L_r}^{a}\right)}\right]^T \in \mathbb{C}^{L_r \times 1}, \label{4}
\end{equation} 
where $2\pi \chi\left(\widehat{\boldsymbol{r}}_n, \varphi_j^e, \varphi_j^a\right)/\lambda$, 
for $1\leq j\leq L_r$, denotes the phase difference of the $j$-th receive path between the position $\widehat{\boldsymbol{r}}_n$ and reference origin of the RIS node. Accordingly, the FRM at RIS can be given as
\begin{equation}
	\boldsymbol{F^r}(\widehat{\boldsymbol{r}}) \triangleq\left[\boldsymbol{f^r}(\widehat{\boldsymbol{r}}_1), \boldsymbol{f^r}(\widehat{\boldsymbol{r}}_2), \ldots, \boldsymbol{f^r}(\widehat{\boldsymbol{r}}_N)\right] \in \mathbb{C}^{L_r \times N}. \label{5}
\end{equation}

Next, let $\boldsymbol{\Lambda} \in \mathbb{C}^{L_r \times L_t}$ denotes the path response matrix for BS-RIS link, then the equivalent channel from BS to the RIS can be written as
\begin{equation}
	\mathbf{H}_{b,r}(\widehat{\boldsymbol{t}})=\boldsymbol{F^r}(\widehat{\boldsymbol{r}})^H \boldsymbol{\Lambda} \boldsymbol{F^t}(\widehat{\boldsymbol{t}}) \in\mathbb{C}^{N \times L},  \label{6}
\end{equation}

Let $\mathbf h_{r,m}\in\mathbb C^{N\times 1}$ denote the estimated RIS--user channel. The estimated effective channel from the BS to user $m$ is expressed as
\begin{equation}
	\mathbf g_m^H(\widehat{\mathbf t},\boldsymbol{\Phi})
	=
	\mathbf h_{r,m}^H\boldsymbol{\Phi}
	\mathbf H_{b,r}(\widehat{\mathbf t})
	\in\mathbb C^{1\times L},
	\label{7}
\end{equation}
where $\boldsymbol{\Phi} =\operatorname{diag}\big\{[e^{j\vartheta_1}, e^{j\vartheta_2},\ldots,e^{j\vartheta_N}]^T\big\} \in\mathbb C^{N\times N}$ denotes the RIS reflection matrix, and $\vartheta_n$ is the phase shift of its $n$-th reflecting element.

\noindent \textit{Remark 1 (Practical implementation considerations)}: Implementing MAs requires positioning actuators, control circuitry, and position feedback in addition to the RF components. Mechanical repositioning introduces movement delays that depend on the actuator implementation, displacement, and movement speed \cite{zhu2025tutorial,wu2025globally}. Control overhead includes channel acquisition, position-command delivery, and synchronization of antenna positioning with the beamforming updates. Repositioning also consumes electrical energy in the MAs drivers, which depends on the movement distance, movement speed, and driver power \cite{wu2025globally}. The proposed design is intended for quasi-static or slowly varying propagation conditions, where channel acquisition, computation, and repositioning can be completed while the estimated spatial channel model remains sufficiently accurate. In the proposed framework, the antenna positions are determined computationally before physical repositioning. Once the optimization is completed, the antennas move to the selected positions and remain fixed during the subsequent data-transmission interval. Less frequent repositioning can reduce the associated time and energy overhead. The reported sum-rate characterizes data transmission at the optimized antenna positions. Movement and control overhead, as well as actuator energy consumption, are outside the scope of the proposed optimization framework.

\subsection{Signal Model}
Based on the RSMA transmission protocol, each user’s message is split into a 
private part and a common part. Specifically, the private messages of the $M$ users are independently encoded into $M$ private data streams, denoted by 
$\{\hat{z}_1,\hat{z}_2,\ldots,\hat{z}_M\}$, where $m\in\mathcal{M}=\{1,2,\ldots,M\}$, while the common messages intended for all users are jointly encoded into a single common data stream, denoted by $\hat{z}_c$. Accordingly, the transmitted signal at the BS is given by
\begin{equation}
	\mathbf{x}= \mathbf{p}_c \hat{z}_c + \sum_{m=1}^M \mathbf{p}_m \hat{z}_m, \ \forall m\in \mathcal M, \label{8}
\end{equation}
where $\mathbf{p}_m \in \mathbb{C}^{L\times 1}$ and $\mathbf{p}_c \in \mathbb{C}^{L\times 1}$ denote the transmit beamforming vectors for the private and common data streams, respectively. 

Next, under imperfect CSI, the actual effective channel of user $m$ is modeled as $\widehat{\mathbf g}_m =\mathbf g_m+\Delta\mathbf g_m$ \cite{zhang2023robust}, $\forall m\in\mathcal M$, where $\mathbf g_m$ denotes the estimated effective channel and $\Delta\mathbf g_m$ represents the channel estimation error. For notational simplicity, the dependence of these channel vectors on $\widehat{\mathbf t}$ and $\boldsymbol{\Phi}$ is omitted. Further, assuming $\mathbb{E}[|\hat z_m|^2]=1$, $\forall m\in\mathcal M$, and $\mathbb{E}[|\hat z_c|^2]=1$, the signal received at the $m$-th user can be written as
\begin{equation}
	\widehat y_m
	=
	\widehat{\mathbf g}_m^H\mathbf x+\widehat n_m,
	\qquad \forall m\in\mathcal M,
	\label{9}
\end{equation}
where $\widehat n_m\sim\mathcal{CN}(0,\widehat{\sigma}_m^2)$ denotes the additive white Gaussian noise at user $m$.

Further, let $\widehat{\mathbf G}_m =\widehat{\mathbf g}_m\widehat{\mathbf g}_m^H$ and $\mathbf G_m=\mathbf g_m\mathbf g_m^H$ denote the channel outer-product matrices associated with the actual
and estimated effective channels, respectively. Both matrices are positive semidefinite (PSD) by construction and satisfy $\widehat{\mathbf G}_m=\mathbf G_m+\Delta\mathbf G_m$, where $\Delta\mathbf G_m$ is the Hermitian matrix representing the estimation error. This error is modeled by the spectral-norm bound $\|\Delta\mathbf G_m\|_2\leq\xi_m$ \cite{zhang2023robust,li2022robust}, where $\xi_m=\rho\|\mathbf G_m\|_2$ and $\rho\in[0,1)$ specifies the relative uncertainty level \cite{li2022robust}. While $\rho$ remains fixed during optimization, the radius $\xi_m$ varies with the MA positions and RIS coefficients through the estimated channel matrix $\mathbf G_m$. In practice, $\rho$ can be calibrated from channel-estimation error measurements or estimator simulations by examining the normalized matrix error $\|\widehat{\mathbf G}_m-\mathbf G_m\|_2/\|\mathbf G_m\|_2$ for nonzero channel estimates and choosing $\rho$ large enough to cover the observed normalized errors, with an additional margin for variations in pilot quality and channel dynamics. For sufficiently stable channels, longer pilot sequences at fixed per-symbol pilot power generally improve channel-estimation accuracy, allowing a smaller value of $\rho$ and hence a smaller uncertainty radius \cite{hassibi2003much}. Conversely, user mobility and longer delays between channel estimation and data transmission can increase the difference between the estimated and actual channels, which may require a larger value of $\rho$ \cite{truong2013effects}.

Thus, the achievable rate for decoding the common stream at the $m$-th user is given by
\begin{equation}
	{R}^c_{m}=\log _2\left(1+{{\Gamma}}^c_{m}\right), \ \forall m\in \mathcal M, \label{10}
\end{equation}
where $\Gamma_m^{c}$ denotes the signal-to-interference-plus-noise ratio (SINR), 
which is given as 
\begin{equation}
	\Gamma_m^c
	=
	\frac{
		\mathbf p_c^H
		(\mathbf G_m+\Delta\mathbf G_m)\mathbf p_c
	}{
		\displaystyle\sum_{j=1}^{M}
		\mathbf p_j^H
		(\mathbf G_m+\Delta\mathbf G_m)\mathbf p_j
		+\widehat{\sigma}_m^2
	},
	\quad\forall m\in\mathcal M.
	\label{11}
\end{equation}

Once the common stream has been successfully decoded, its effect is eliminated from the received signal via successive interference cancellation (SIC) \cite{mao2022rate}. The receiver then proceeds to decode its intended private data stream, while treating the private streams intended for other users as residual interference. Accordingly, the achievable rate associated with the private message for the $m$-th user can be expressed as
\begin{equation}
	R^p_{m}=\log _2\left(1+{\Gamma}^p_{m}\right), \ \forall m\in \mathcal M, \label{12}
\end{equation}
where $\Gamma_m^{p}$ represents the SINR, which is given by
\begin{equation}
	\Gamma_m^p
	=
	\frac{
		\mathbf p_m^H
		(\mathbf G_m+\Delta\mathbf G_m)\mathbf p_m
	}{
		\displaystyle\sum_{\substack{j=1\\j\neq m}}^{M}
		\mathbf p_j^H
		(\mathbf G_m+\Delta\mathbf G_m)\mathbf p_j
		+\widehat{\sigma}_m^2
	},
	\quad\forall m\in\mathcal M.
	\label{13}
\end{equation}

Accordingly, the sum-rate of the considered system can be written as

	\begin{align}
	\widehat{R}= & \sum\limits_{{\substack{m \in \mathcal M}}} \Big(\widehat{r}^c_m + {R}^p_{m}\Big), \ \forall m\in \mathcal M,  \label{14}
\end{align}
where $\widehat{r}^c_m$ denotes the actual common rate allocated to $m$-th user.

\subsection{Problem Formulation}
This work focuses on robust sum-rate maximization for a MAs-enabled RSMA-based RIS system by jointly optimizing the transmit precoding vectors $\mathbf p_m$ and $\mathbf p_c$, the common-rate allocation, the RIS reflection matrix $\boldsymbol{\Phi}$, and the MA positions $\widehat{\boldsymbol t}$, subject to QoS requirements under a bounded CSI error model. Let $\underline R_m^s =\min_{\Delta\mathbf G_m\in\mathcal U_m}R_m^s$, $s\in\{c,p\}$, denote the worst-case common- and
private-stream rates of user $m$, respectively, where $\mathcal U_m =\{\Delta\mathbf G_m=\Delta\mathbf G_m^H: \|\Delta\mathbf G_m\|_2\leq\xi_m,\, \mathbf G_m+\Delta\mathbf G_m\succeq\mathbf0\}$ denotes the admissible matrix-error set. Accordingly, the resulting optimization problem is formulated as follows:
\begin{subequations}\label{P1}
	\begin{align}
		\text{\textbf{(P1)}}\quad
		&\max_{\{\widehat r_m^c\},\,\mathbf p_c,\,
			\{\mathbf p_m\},\,\boldsymbol{\Phi},\,
			\widehat{\boldsymbol t}}
		\sum_{m=1}^{M}
		\left(\widehat r_m^c+\underline R_m^p\right),
		\label{15a}\\
		\mathrm{s.t.}\quad C_1:\;&
		\widehat r_m^c+\underline R_m^p\geq R_{\min},
		\quad\forall m\in\mathcal M,
		\label{15b}\\
		C_2:\;&
		\sum_{k=1}^{M}\widehat r_k^c
		\leq\underline R_m^c,
		\quad\forall m\in\mathcal M,
		\label{15c}\\
		C_3:\;&
		\|\mathbf p_c\|_2^2
		+\sum_{m=1}^{M}\|\mathbf p_m\|_2^2
		\leq P_t,
		\label{15d}\\
		C_4:\;&
		|e^{j\vartheta_i}|=1,
		\quad\forall i\in\mathcal N,
		\label{15e}\\
		C_5:\;&
		\vartheta_i\in[0,2\pi],
		\quad\forall i\in\mathcal N,
		\label{15f}\\
		C_6:\;&
		\|\widehat{\boldsymbol t}_l
		-\widehat{\boldsymbol t}_j\|_2\geq D,
		\quad
		\forall l,j\in\mathcal L,\quad l\neq j,
		\label{15g}\\
		C_7:\;&
		\widehat{\boldsymbol t}_l\in\mathcal C_t,
		\quad\forall l\in\mathcal L,
		\label{15h}\\
		C_8:\;&
		\widehat r_m^c\geq0,
		\quad\forall m\in\mathcal M.
		\label{15i}
	\end{align}
\end{subequations}
where $P_t$, $R_{\min}$, and $D$ denote the available power budget, the minimum rate threshold desired for QoS assurance, and the minimum separation distance between MAs to avoid mutual coupling effects, respectively. Constraint $C_1$ is imposed to satisfy QoS requirements, whereas $C_2$ guarantees successful decoding of the common stream. Constraint $C_4$ enforces the unit-modulus condition on the reflecting elements of the RIS. Constraint $C_6$ enforces a minimum separation $D$ between MAs to mitigate mutual coupling \cite{zhu2025tutorial}. Residual coupling can alter the effective channels and change the desired-signal and interference powers achieved by the optimized precoders. If the total channel mismatch, measured by $\|\widehat{\mathbf G}_m-\mathbf G_m\|_2$, exceeds the adopted uncertainty radius $\xi_m$, the achieved rates of user $m$ may fall below the conservative rate bounds used in the design. Constraint $C_7$ restricts the MA positions to lie within the region $\mathcal{C}_t$. Constraint $C_8$ ensures nonnegative common-rate allocations.

\section{Proposed Solution for the Optimization Problem}
 The coupling among the transmit precoders, RIS reflection coefficients, and MAs positions in the rate expressions renders \textbf{(P1)} non-convex. We therefore decompose the problem through alternating optimization and construct tractable approximations for the resulting subproblems. Each stage updates one group of design variables while holding the other groups at their latest values. In particular, the precoding update uses the estimated effective channels and uncertainty radii associated with the current RIS coefficients and MAs positions, while the RIS and MAs position updates evaluate the conservative rates using the current precoders. All updates pursue the same conservative sum-rate objective, with the common-rate allocation reoptimized at each stage.
\subsection{Updating Precoding Vectors $\mathbf{p}_{c}$ and  $\mathbf{p}_{m}$}
With the RIS reflection matrix $\boldsymbol{\Phi}$ and MA positions $\widehat{\boldsymbol t}$ fixed, the estimated channel matrices $\mathbf G_m$ and uncertainty radii $\xi_m$ remain constant. The precoding vectors and common-rate allocations are optimized through
\begin{subequations}\label{P2}
	\begin{align}
		\text{\textbf{(P2)}}\quad
		&\max_{\{\widehat r_m^c\},\,\mathbf p_c,\,
			\{\mathbf p_m\}}
		\sum_{m=1}^{M}
		\left(\widehat r_m^c+\underline R_m^p\right),
		\label{16a}\\
		\mathrm{s.t.}\quad
		&\eqref{15b},\eqref{15c},\eqref{15d},\eqref{15i}.
		\label{16b}
	\end{align}
\end{subequations}

It should be noted that Problem \textbf{(P2)} is non-convex due to the coupled precoding variables in the worst-case rate expressions.

Next, we define $\mathbf P_c=\mathbf p_c\mathbf p_c^H$ and $\mathbf P_m=\mathbf p_m\mathbf p_m^H$ as PSD matrices satisfying $\operatorname{rank}(\mathbf P_c)=1$ and $\operatorname{rank}(\mathbf P_m)=1$, respectively. To bound the uncertain received-power terms, we use the following result.

\noindent\textbf{Theorem 1:}
Let $\widehat{\mathbf\Psi}$ and $\widehat{\mathbf\Xi}$ be Hermitian matrices, with $\widehat{\mathbf\Psi}\succeq\mathbf0$. For a given $\xi_m\geq0$,
\begin{equation}
	\max_{\|\widehat{\mathbf\Xi}\|_2\leq\xi_m}
	\operatorname{tr}
	(\widehat{\mathbf\Psi}\widehat{\mathbf\Xi})
	=
	\xi_m\operatorname{tr}(\widehat{\mathbf\Psi}).
	\label{17}
\end{equation}
\noindent\textbf{Proof:} The proof is provided in Appendix~A.

Let $\mathbf S^c=\sum_{j=1}^{M}\mathbf P_j$ and $\mathbf S_m^p=\sum_{j\neq m}\mathbf P_j$ denote the interference covariance matrices for common- and private-stream decoding, respectively. Using \textbf{Theorem~1}, we lower-bound the desired-signal power and upper-bound the interference power. The resulting conservative common- and private-stream rate are given as
\begin{align}
	\overline R_m^c
	&=
	\log_2\left(
	1+
	\frac{
		\operatorname{tr}
		[(\mathbf G_m-\xi_m\mathbf I_L)\mathbf P_c]
	}{
		\operatorname{tr}
		[(\mathbf G_m+\xi_m\mathbf I_L)\mathbf S^c]
		+\widehat{\sigma}_m^2
	}
	\right),
	\label{18}\\
	\overline R_m^p
	&=
	\log_2\left(
	1+
	\frac{
		\operatorname{tr}
		[(\mathbf G_m-\xi_m\mathbf I_L)\mathbf P_m]
	}{
		\operatorname{tr}
		[(\mathbf G_m+\xi_m\mathbf I_L)\mathbf S_m^p]
		+\widehat{\sigma}_m^2
	}
	\right),
	\label{19}
\end{align}
for all $m\in\mathcal M$. To ensure nonnegative desired-signal lower bounds,
we impose
\begin{subequations}\label{AB_signal_nonnegative}
	\begin{align}
		\operatorname{tr}
		[(\mathbf G_m-\xi_m\mathbf I_L)\mathbf P_c]
		&\geq0,
		\quad\forall m\in\mathcal M,
		\label{20a}\\
		\operatorname{tr}
		[(\mathbf G_m-\xi_m\mathbf I_L)\mathbf P_m]
		&\geq0,
		\quad\forall m\in\mathcal M.
		\label{20b}
	\end{align}
\end{subequations}
Under these conditions, $\overline R_m^s\leq R_m^s$ for every $\Delta\mathbf G_m\in\mathcal U_m$,
$s\in\{c,p\}$. Consequently, $\overline R_m^s\leq\underline R_m^s$, $\forall m\in\mathcal M$, $s\in\{c,p\}$.

Further, using $\overline R_m^c$ and $\overline R_m^p$,
a conservative approximation of\textbf{ (P2)} is formulated as follows:
\begin{subequations}\label{P3}
	\begin{align}
		\text{\textbf{(P3)}}\quad
		&\max_{\mathbf P_c,\{\mathbf P_m,\widehat r_m^c\}}
		\sum_{m=1}^{M}
		\left(\widehat r_m^c+\overline R_m^p\right),
		\label{21a}\\
		\mathrm{s.t.}\quad
		&\widehat r_m^c+\overline R_m^p
		\geq R_{\min},
		\quad\forall m\in\mathcal M,
		\label{21b}\\
		&\sum_{k=1}^{M}\widehat r_k^c
		\leq\overline R_m^c,
		\quad\forall m\in\mathcal M,
		\label{21c}\\
		&\operatorname{tr}(\mathbf P_c)
		+\sum_{m=1}^{M}\operatorname{tr}(\mathbf P_m)
		\leq P_t,
		\label{21d}\\
		&\mathbf P_c\succeq\mathbf0,\quad
		\mathbf P_m\succeq\mathbf0,
		\quad\forall m\in\mathcal M,
		\label{21e}\\
		&\operatorname{rank}(\mathbf P_c)=1,\quad
		\operatorname{rank}(\mathbf P_m)=1,
		\quad\forall m\in\mathcal M,
		\label{21f}\\
		&\eqref{15i},\eqref{20a},
		\eqref{20b}.
		\label{21g}
	\end{align}
\end{subequations}

To facilitate the convex approximation of \textbf{(P3)}, introduce the slack-variable vector
$\boldsymbol{\kappa}^p =[\kappa_1^p,\ldots,\kappa_M^p]^T$, where $\kappa_m^p$ lower-bounds the conservative private-stream rate. Accordingly, we impose
\begin{subequations}\label{22}
	\begin{align}
		\widehat r_m^c+\kappa_m^p
		&\geq R_{\min},
		\quad\forall m\in\mathcal M,
		\label{22a}\\
		\overline R_m^p
		&\geq\kappa_m^p,
		\quad\forall m\in\mathcal M.
		\label{22b}
	\end{align}
\end{subequations}

To handle \eqref{22b}, introduce the auxiliary-variable vector $\boldsymbol{\zeta}^p =[\zeta_1^p,\ldots,\zeta_M^p]^T$, where $\zeta_m^p>0$ upper-bounds the conservative interference-plus-noise term. Constraint \eqref{22b} can then be represented by
\begin{subequations}\label{23}
	\begin{align}
		&\operatorname{tr}
		[(\mathbf G_m-\xi_m\mathbf I_L)\mathbf P_m]
		\geq(2^{\kappa_m^p}-1)\zeta_m^p,
		\nonumber\\
		&\qquad\forall m\in\mathcal M,
		\label{23a}\\
		&\operatorname{tr}
		[(\mathbf G_m+\xi_m\mathbf I_L)\mathbf S_m^p]
		+\widehat{\sigma}_m^2
		\leq\zeta_m^p,
		\nonumber\\
		&\qquad\forall m\in\mathcal M.
		\label{23b}
	\end{align}
\end{subequations}

Constraint \eqref{23a} remains non-convex due to the coupled term $(2^{\kappa_m^p}-1)\zeta_m^p$.
We introduce $\gamma_m^p\geq0$ satisfying
\begin{equation}
	2^{\kappa_m^p}-1\leq\gamma_m^p,
	\qquad\forall m\in\mathcal M.
	\label{24}
\end{equation}
At iteration $i$, define $d_m^{(i)}=\gamma_m^{p,(i)}-\zeta_m^{p,(i)}$. Using the difference-of-squares identity, a convex upper bound on $\gamma_m^p\zeta_m^p$ is obtained as
\begin{align}
	\gamma_m^p\zeta_m^p
	&\leq\mathcal Q_m^{(i)}(\gamma_m^p,\zeta_m^p)
	\nonumber\\
	&\triangleq\frac14\Big[
	(\gamma_m^p+\zeta_m^p)^2
	-2d_m^{(i)}(\gamma_m^p-\zeta_m^p)
	+(d_m^{(i)})^2
	\Big].
	\label{25}
\end{align}
This upper bound is exact at the current iterate. Accordingly, a convex sufficient approximation of \eqref{23a}
is therefore given by \eqref{24}, $\gamma_m^p\geq0$, and
\begin{align}
	&\mathcal Q_m^{(i)}(\gamma_m^p,\zeta_m^p)
	\leq
	\operatorname{tr}
	[(\mathbf G_m-\xi_m\mathbf I_L)\mathbf P_m],
	\nonumber\\
	&\qquad\forall m\in\mathcal M.
	\label{26}
\end{align}

To handle \eqref{21c}, introduce the auxiliary-variable vector $\boldsymbol{\beta}^c =[\beta_1^c,\ldots,\beta_M^c]^T$, where $\beta_m^c>0$ upper-bounds the conservative interference-plus-noise term during common-stream decoding. Constraint \eqref{21c} can then be represented by
\begin{subequations}\label{27}
	\begin{align}
		&\operatorname{tr}
		[(\mathbf G_m-\xi_m\mathbf I_L)\mathbf P_c]
		\geq
		\left(2^{\sum_{k=1}^{M}\widehat r_k^c}-1\right)
		\beta_m^c,
		\nonumber\\
		&\qquad\forall m\in\mathcal M,
		\label{27a}\\
		&\operatorname{tr}
		[(\mathbf G_m+\xi_m\mathbf I_L)\mathbf S^c]
		+\widehat{\sigma}_m^2
		\leq\beta_m^c,
		\nonumber\\
		&\qquad\forall m\in\mathcal M.
		\label{27b}
	\end{align}
\end{subequations}

To address the non-convex product in \eqref{27a}, introduce $\gamma_m^c\geq0$ satisfying
\begin{equation}
	2^{\sum_{k=1}^{M}\widehat r_k^c}-1
	\leq\gamma_m^c,
	\qquad\forall m\in\mathcal M.
	\label{28}
\end{equation}
At iteration $i$, define $d_{m,c}^{(i)} =\gamma_m^{c,(i)}-\beta_m^{c,(i)}$, a convex
upper bound on $\gamma_m^c\beta_m^c$ is
\begin{align}
	\gamma_m^c\beta_m^c
	&\leq
	\mathcal Q_{m,c}^{(i)}(\gamma_m^c,\beta_m^c)
	\nonumber\\
	&\triangleq
	\frac14\Big[
	(\gamma_m^c+\beta_m^c)^2
	-2d_{m,c}^{(i)}(\gamma_m^c-\beta_m^c)
	+(d_{m,c}^{(i)})^2
	\Big].
	\label{29}
\end{align}
This upper bound is exact at the current iterate. A convex sufficient approximation of \eqref{27a}
is therefore given by \eqref{28}, $\gamma_m^c\geq0$, and
\begin{align}
	&\mathcal Q_{m,c}^{(i)}(\gamma_m^c,\beta_m^c)
	\leq
	\operatorname{tr}
	[(\mathbf G_m-\xi_m\mathbf I_L)\mathbf P_c],
	\nonumber\\
	&\qquad\forall m\in\mathcal M.
	\label{30}
\end{align}

Accordingly, the precoding subproblem is approximated as
\begin{subequations}\label{P4}
	\begin{align}
		\text{\textbf{(P4)}}\quad
		&\max_{\widehat{\boldsymbol{\alpha}}}
		\sum_{m=1}^{M}
		\left(\widehat r_m^c+\kappa_m^p\right),
		\label{31a}\\
		\mathrm{s.t.}\quad
		&\eqref{22a},\eqref{23b},\eqref{24},\eqref{26},
		\eqref{27b},\eqref{28},\eqref{30},
		\label{31b}\\
		&\eqref{21d},\eqref{21e},\eqref{21f},
		\eqref{21g},
		\label{31c}\\
        &\kappa_m^p\geq0,\quad
        \gamma_m^p\geq0,\quad
        \gamma_m^c\geq0,
        \quad\forall m\in\mathcal M,
        \label{31d}
	\end{align}
\end{subequations}
where $\widehat{\boldsymbol{\alpha}}$ collects $\mathbf P_c$ and $\{\mathbf P_m,\widehat r_m^c,\kappa_m^p,\zeta_m^p,\beta_m^c,\gamma_m^p,\gamma_m^c\}_{m\in\mathcal M}$. Problem \textbf{(P4)} remains non-convex due to the rank-one constraints in \eqref{21f}. Relaxing these constraints yields a convex problem for fixed $d_m^{(i)}$ and $d_{m,c}^{(i)}$. If all relaxed precoding matrices are rank one, the precoding vectors are recovered from their principal eigenpairs. Otherwise, we perform $J_{\mathrm p}$ Gaussian randomization trials \cite{ni2021resource}. In each trial, the common and private precoding vectors are drawn from zero-mean circularly symmetric complex Gaussian distributions with the corresponding relaxed matrices as their covariances. The sampled vectors together form one candidate precoding design. Each design is checked against the constraints of \textbf{(P3)} using its associated common-rate allocation. We select the design with the largest conservative sum-rate among the feasible candidates and the previous feasible design, retaining the previous design if no candidate achieves a strictly higher value.

\subsection{Updating RIS Reflection Matrix $\mathbf \Phi$}
With the precoding vectors and MA positions fixed, the RIS reflection matrix and common-rate allocations are optimized through
\begin{subequations}\label{P5}
	\begin{align}
		\text{\textbf{(P5)}}\quad
		&\max_{\boldsymbol{\Phi},\{\widehat r_m^c\}}
		\sum_{m=1}^{M}
		\left(\widehat r_m^c+\overline R_m^p\right),
		\label{32a}\\
		\mathrm{s.t.}\quad
		&\widehat r_m^c+\overline R_m^p
		\geq R_{\min},
		\quad\forall m\in\mathcal M,
		\label{32b}\\
		&\sum_{k=1}^{M}\widehat r_k^c
		\leq\overline R_m^c,
		\quad\forall m\in\mathcal M,
		\label{32c}\\
		&\eqref{15e},\eqref{15f},\eqref{15i},\eqref{20a},\eqref{20b}.
		\label{32d}
	\end{align}
\end{subequations}
Although the precoding matrices remain fixed, $\mathbf G_m$ and the uncertainty radius $\xi_m=\rho\|\mathbf G_m\|_2$ vary with $\boldsymbol{\Phi}$. This dependence is explicitly incorporated into the following reformulation.

Next, define $\mathbf v
=[e^{j\vartheta_1},\ldots,e^{j\vartheta_N}]^T$
and $\widehat{\mathbf V}=\mathbf v\mathbf v^H$,
where $\widehat{\mathbf V}\succeq\mathbf0$,
$\operatorname{rank}(\widehat{\mathbf V})=1$,
and $[\widehat{\mathbf V}]_{n,n}=1$,
$\forall n\in\mathcal N$.
Let
$\mathbf D_m
=\operatorname{diag}(\mathbf h_{r,m}^H)\mathbf H_{b,r}
\in\mathbb C^{N\times L}$.
Since $\mathbf g_m^H=\mathbf v^T\mathbf D_m$, the uncertainty radius is expressed as
\begin{align}
	\xi_m(\widehat{\mathbf V})
	&=\rho\|\mathbf G_m\|_2
	=\rho\|\mathbf g_m\|_2^2
	\nonumber\\
	&=\rho\operatorname{tr}
	\left[
	(\mathbf D_m\mathbf D_m^H)^T
	\widehat{\mathbf V}
	\right],
	\quad\forall m\in\mathcal M.
	\label{33}
\end{align}

Using \eqref{33}, the conservative rate bounds in \eqref{18} and \eqref{19} can be rewritten as
\begin{align}
	\overline R_m^c(\widehat{\mathbf V})
	&=
	\log_2\left(
	1+
	\frac{
		\operatorname{tr}
		(\widehat{\mathbf V}\widehat{\mathbf U}_{m,1})
	}{
		\operatorname{tr}
		(\widehat{\mathbf V}\widehat{\mathbf U}_{m,2})
		+\widehat{\sigma}_m^2
	}
	\right),
	\nonumber\\
	&\qquad\forall m\in\mathcal M,
	\label{34}\\
	\overline R_m^p(\widehat{\mathbf V})
	&=
	\log_2\left(
	1+
	\frac{
		\operatorname{tr}
		(\widehat{\mathbf V}\widehat{\mathbf W}_{m,1})
	}{
		\operatorname{tr}
		(\widehat{\mathbf V}\widehat{\mathbf W}_{m,2})
		+\widehat{\sigma}_m^2
	}
	\right),
	\nonumber\\
	&\qquad\forall m\in\mathcal M,
	\label{35}
\end{align}
where
\begin{align*}
	\widehat{\mathbf U}_{m,1}
	&=
	\left(
	\mathbf D_m
	[\mathbf P_c
	-\rho\operatorname{tr}(\mathbf P_c)\mathbf I_L]
	\mathbf D_m^H
	\right)^T,
	\\
	\widehat{\mathbf U}_{m,2}
	&=
	\left(
	\mathbf D_m
	[\mathbf S^c
	+\rho\operatorname{tr}(\mathbf S^c)\mathbf I_L]
	\mathbf D_m^H
	\right)^T,
	\\
	\widehat{\mathbf W}_{m,1}
	&=
	\left(
	\mathbf D_m
	[\mathbf P_m
	-\rho\operatorname{tr}(\mathbf P_m)\mathbf I_L]
	\mathbf D_m^H
	\right)^T,
	\\
	\widehat{\mathbf W}_{m,2}
	&=
	\left(
	\mathbf D_m
	[\mathbf S_m^p
	+\rho\operatorname{tr}(\mathbf S_m^p)\mathbf I_L]
	\mathbf D_m^H
	\right)^T.
\end{align*}
These coefficient matrices remain fixed during the
passive-beamforming update.

Further, using \eqref{34} and \eqref{35}, problem \textbf{(P5)}
can be equivalently expressed as
\begin{subequations}\label{P6}
	\begin{align}
		\text{\textbf{(P6)}}\quad
		&\max_{\widehat{\mathbf V},\{\widehat r_m^c\}}
		\sum_{m=1}^{M}
		\left(\widehat r_m^c+\overline R_m^p(\widehat{\mathbf V})\right),
		\label{36a}\\
		\mathrm{s.t.}\quad
		&\widehat r_m^c+\overline R_m^p(\widehat{\mathbf V})
		\geq R_{\min},
		\quad\forall m\in\mathcal M,
		\label{36b}\\
		&\sum_{k=1}^{M}\widehat r_k^c
		\leq \overline R_m^c(\widehat{\mathbf V}),
		\quad\forall m\in\mathcal M,
		\label{36c}\\
		&[\widehat{\mathbf V}]_{n,n}=1,
		\quad\forall n\in\mathcal N,
		\label{36d}\\
		&\widehat{\mathbf V}\succeq\mathbf0,
		\label{36e}\\
		&\operatorname{rank}(\widehat{\mathbf V})=1,
		\label{36f}
		\end{align}
		\begin{align}
		&\operatorname{tr}
		(\widehat{\mathbf V}\widehat{\mathbf U}_{m,1})
		\geq0, \;
	     \operatorname{tr}
		(\widehat{\mathbf V}\widehat{\mathbf W}_{m,1})
		\geq0, 	\nonumber\\
		&\quad\forall m\in\mathcal M.
		\label{36g}\\
		&\widehat r_m^c\geq0,
		\quad\forall m\in\mathcal M,
		\label{36h}
	\end{align}
\end{subequations}

To handle the objective and constraint \eqref{36b}, introduce the slack-variable vector $\boldsymbol{\varpi}^p =[\varpi_1^p,\ldots,\varpi_M^p]^T$ satisfying
\begin{subequations}\label{37}
	\begin{align}
		\widehat r_m^c
		+\overline R_m^p(\widehat{\mathbf V})
		&\geq\varpi_m^p,
		\quad\forall m\in\mathcal M,
		\label{37a}\\
		\varpi_m^p&\geq R_{\min},
		\quad\forall m\in\mathcal M.
		\label{37b}
	\end{align}
\end{subequations}
The corresponding objective is $\sum_{m=1}^{M}\varpi_m^p$. Using \eqref{35}, constraint \eqref{37a} becomes
\begin{align}
	&\widehat r_m^c
	+\log_2\left(
	\operatorname{tr}
	[\widehat{\mathbf V}
	(\widehat{\mathbf W}_{m,1}
	+\widehat{\mathbf W}_{m,2})]
	+\widehat{\sigma}_m^2
	\right)
	\nonumber\\
	&\quad
	-\log_2\left(
	\operatorname{tr}
	(\widehat{\mathbf V}\widehat{\mathbf W}_{m,2})
	+\widehat{\sigma}_m^2
	\right)
	\geq\varpi_m^p,
	\quad\forall m\in\mathcal M.
	\label{38}
\end{align}

At iteration $i$, the concavity of the logarithm gives the following affine upper bound:
\begin{align}
	&\log_2\left(
	\operatorname{tr}
	(\widehat{\mathbf V}\widehat{\mathbf W}_{m,2})
	+\widehat{\sigma}_m^2
	\right)
	\leq\Pi_m^{(i)}(\widehat{\mathbf V}),
	\label{39}
\end{align}
where
\begin{align}
	\Pi_m^{(i)}(\widehat{\mathbf V})
	\triangleq{}&
	\log_2\left(
	\operatorname{tr}
	(\widehat{\mathbf V}^{(i)}
	\widehat{\mathbf W}_{m,2})
	+\widehat{\sigma}_m^2
	\right)
	\nonumber\\
	&+
	\frac{
		\operatorname{tr}\left[
		(\widehat{\mathbf V}-\widehat{\mathbf V}^{(i)})
		\widehat{\mathbf W}_{m,2}
		\right]
	}{
		\left(
		\operatorname{tr}
		(\widehat{\mathbf V}^{(i)}
		\widehat{\mathbf W}_{m,2})
		+\widehat{\sigma}_m^2
		\right)\ln 2
	}.
	\label{40}
\end{align}
Here, $\widehat{\mathbf V}^{(i)}$ is the current iterate. Replacing the second logarithm in \eqref{38}
with this upper bound yields the convex sufficient constraint
\begin{align}
	&\widehat r_m^c
	+\log_2\left(
	\operatorname{tr}
	[\widehat{\mathbf V}
	(\widehat{\mathbf W}_{m,1}
	+\widehat{\mathbf W}_{m,2})]
	+\widehat{\sigma}_m^2
	\right)
	\nonumber\\
	&\quad
	-\Pi_m^{(i)}(\widehat{\mathbf V})
	\geq\varpi_m^p,
	\quad\forall m\in\mathcal M.
	\label{41}
\end{align}

Similarly, using \eqref{34}, constraint \eqref{36c} can be written as
\begin{align}
	&\log_2\left(
	\operatorname{tr}
	[\widehat{\mathbf V}
	(\widehat{\mathbf U}_{m,1}
	+\widehat{\mathbf U}_{m,2})]
	+\widehat{\sigma}_m^2
	\right)
	\nonumber\\
	&\quad
	-\log_2\left(
	\operatorname{tr}
	(\widehat{\mathbf V}\widehat{\mathbf U}_{m,2})
	+\widehat{\sigma}_m^2
	\right)
	\geq\sum_{k=1}^{M}\widehat r_k^c,
	\nonumber\\
	&\qquad\forall m\in\mathcal M.
	\label{42}
\end{align}
An affine upper bound on the second logarithm is
\begin{align}
	\Pi_{m,c}^{(i)}(\widehat{\mathbf V})
	\triangleq{}&
	\log_2\left(
	\operatorname{tr}
	(\widehat{\mathbf V}^{(i)}
	\widehat{\mathbf U}_{m,2})
	+\widehat{\sigma}_m^2
	\right)
	\nonumber\\
	&+
	\frac{
		\operatorname{tr}\left[
		(\widehat{\mathbf V}-\widehat{\mathbf V}^{(i)})
		\widehat{\mathbf U}_{m,2}
		\right]
	}{
		\left(
		\operatorname{tr}
		(\widehat{\mathbf V}^{(i)}
		\widehat{\mathbf U}_{m,2})
		+\widehat{\sigma}_m^2
		\right)\ln 2
	}.
	\label{43}
\end{align}
Consequently, a convex sufficient approximation of \eqref{42} is
\begin{align}
	&\log_2\left(
	\operatorname{tr}
	[\widehat{\mathbf V}
	(\widehat{\mathbf U}_{m,1}
	+\widehat{\mathbf U}_{m,2})]
	+\widehat{\sigma}_m^2
	\right)
	\nonumber\\
	&\quad
	-\Pi_{m,c}^{(i)}(\widehat{\mathbf V})
	\geq\sum_{k=1}^{M}\widehat r_k^c,
	\quad\forall m\in\mathcal M.
	\label{44}
\end{align}

Accordingly, the passive-beamforming subproblem is approximated as
\begin{subequations}\label{P7}
	\begin{align}
		\text{\textbf{(P7)}}\quad
		&\max_{\widehat{\mathbf V},
			\{\widehat r_m^c,\varpi_m^p\}}
		\sum_{m=1}^{M}\varpi_m^p,
		\label{45a}\\
		\mathrm{s.t.}\quad
		&\varpi_m^p\geq R_{\min},
		\quad\forall m\in\mathcal M,
		\label{45b}\\
		&\eqref{41},\eqref{44},
		\label{45c}\\
		&\eqref{36d},\eqref{36e},\eqref{36f},
		\eqref{36g},\eqref{36h}.
		\label{45d}
	\end{align}
\end{subequations}
Problem \textbf{(P7)} remains non-convex due to the rank-one constraint in \eqref{36f}. Relaxing this constraint yields a convex problem for fixed $\widehat{\mathbf V}^{(i)}$. If the relaxed RIS matrix is rank one, the reflection vector is recovered from its principal eigenpair. Otherwise, we perform $J_{\mathrm r}$ Gaussian randomization trials. In each trial, a vector is drawn from a zero-mean circularly symmetric complex Gaussian distribution with the relaxed RIS matrix as its covariance. Each entry is then replaced by a unit-modulus coefficient with the same phase. For each candidate, the effective channels and uncertainty radii are recomputed, and the candidate is checked against the constraints of \textbf{(P5)} with its associated common-rate allocation. We select the design with the largest conservative sum-rate among the feasible candidates and the current feasible design, retaining the current design if no candidate achieves a strictly higher value.

\subsection{Updating MAs Positions $\widehat{\mathbf t}$}
After updating the precoding vectors and RIS reflection coefficients, the positions of the BS MAs are optimized together with the common-rate allocation. The conservative rates depend on the antenna positions through both $\mathbf G_m(\widehat{\boldsymbol t})$ and the uncertainty radius $\xi_m(\widehat{\boldsymbol t})=\rho\|\mathbf G_m(\widehat{\boldsymbol t})\|_2$. The subproblem for optimizing the MA positions is therefore formulated as follows:
\begin{subequations}\label{P8}
	\begin{align}
		\text{\textbf{(P8)}}\quad
		&\max_{\widehat{\boldsymbol t},
			\{\widehat r_m^c\}}
		\sum_{m=1}^{M}
		\left(
		\widehat r_m^c+
		\overline R_m^p(\widehat{\boldsymbol t})
		\right),
		\label{46a}\\
		\mathrm{s.t.}\quad
		&\widehat r_m^c+
		\overline R_m^p(\widehat{\boldsymbol t})
		\geq R_{\min},
		\qquad\forall m\in\mathcal M,
		\label{46b}\\
		&\sum_{k=1}^{M}\widehat r_k^c
		\leq\overline R_m^c(\widehat{\boldsymbol t}),
		\quad\forall m\in\mathcal M,
		\label{46c}\\
		&\eqref{15g},\eqref{15h},\eqref{15i},
		\eqref{20a},\eqref{20b}.
		\label{46d}
	\end{align}
\end{subequations}
Problem \textbf{(P8)} is non-convex because the effective channels depend non-linearly on the MA positions and the minimum antenna-separation constraints couple their locations. We address this difficulty by adjusting one antenna position at a time, keeping the remaining positions fixed, and constructing a convex approximation around its current feasible position.

When updating antenna $l$, let $\mathbf u=\widehat{\boldsymbol t}_l$ denote its position variable and $\mathbf u^{(i)}$ its current position. Define $\Delta\mathbf u=\mathbf u-\mathbf u^{(i)}$ as the displacement from the current position. Throughout this update, the positions of all other antennas remain fixed. The search is restricted to a compact convex region $\mathcal C_l^{(i)} \subseteq\mathcal C_t$ containing $\mathbf u^{(i)}$. This region is selected so that the rate denominators and logarithm arguments remain strictly positive, ensuring that the rate functions are twice continuously differentiable on a neighborhood of the region. For notational simplicity, write the estimated channel and its outer-product matrix as $\mathbf g_m(\mathbf u)$ and $\mathbf G_m(\mathbf u)$, respectively. The uncertainty radius is evaluated as $\xi_m(\mathbf u)=\rho\|\mathbf G_m(\mathbf u)\|_2=\rho\|\mathbf g_m(\mathbf u)\|_2^2$, $\forall m\in\mathcal M$. Its dependence on $\mathbf u$ is retained when computing the rate gradients and curvature bounds.

The following lemma provides the lower approximation used in the position update.

\noindent\textbf{Lemma 1:}
Let $f:\mathbb R^2\rightarrow\mathbb R$ be twice continuously differentiable on a neighborhood of
$\mathcal C_l^{(i)}$. Suppose that a constant $\chi_f^{(i)}\geq0$ satisfies
\begin{equation}
	\nabla^2 f(\mathbf u)
	\succeq-\chi_f^{(i)}\mathbf I_2,
	\qquad\forall\mathbf u\in\mathcal C_l^{(i)}.
	\label{47}
\end{equation}
Then, for every $\mathbf u\in\mathcal C_l^{(i)}$,
\begin{align}
	f(\mathbf u)
	&\geq
	f(\mathbf u^{(i)})
	+\nabla f(\mathbf u^{(i)})^T\Delta\mathbf u
	\nonumber\\
	&\quad
	-\frac{\chi_f^{(i)}}{2}
	\|\Delta\mathbf u\|_2^2
	\triangleq\underline f^{(i)}(\mathbf u).
	\label{48}
\end{align}
The function $\underline f^{(i)}$ is concave and matches both the value and gradient of $f$ at $\mathbf u^{(i)}$.

\noindent\textbf{Proof:} The proof is provided in Appendix~B.

Applying \textbf{Lemma~1} to $\overline R_m^c(\mathbf u)$ and $\overline R_m^p(\mathbf u)$ gives the concave quadratic lower bounds
\begin{subequations}\label{49}
	\begin{align}
		\widetilde R_m^{c,(i)}(\mathbf u)
		&=
		\overline R_m^c(\mathbf u^{(i)})
		+\nabla\overline R_m^c(\mathbf u^{(i)})^T
		\Delta\mathbf u
		\nonumber\\
		&\quad
		-\frac{\chi_m^{c,(i)}}{2}
		\|\Delta\mathbf u\|_2^2,
		\label{49a}\\
		\widetilde R_m^{p,(i)}(\mathbf u)
		&=
		\overline R_m^p(\mathbf u^{(i)})
		+\nabla\overline R_m^p(\mathbf u^{(i)})^T
		\Delta\mathbf u
		\nonumber\\
		&\quad
		-\frac{\chi_m^{p,(i)}}{2}
		\|\Delta\mathbf u\|_2^2,
		\label{49b}
	\end{align}
\end{subequations}
where $\chi_m^{s,(i)}\geq0$, $s\in\{c,p\}$, satisfies
\begin{equation}
	\nabla^2\overline R_m^s(\mathbf u)
	\succeq-\chi_m^{s,(i)}\mathbf I_2,
	\quad
	\forall\mathbf u\in\mathcal C_l^{(i)}.
	\label{50}
\end{equation}
Consequently, $\widetilde R_m^{s,(i)}(\mathbf u)\leq\overline R_m^s(\mathbf u)$ throughout $\mathcal C_l^{(i)}$. Both bounds match the corresponding rate values and gradients at $\mathbf u^{(i)}$. The gradients and curvature bounds account for the position dependence of both $\mathbf G_m(\mathbf u)$
and $\xi_m(\mathbf u)$.

To handle \eqref{20a} and \eqref{20b}, we define the desired-signal lower-bound functions as
\begin{subequations}\label{51}
	\begin{align}
		A_m^c(\mathbf u)
		&=
		\operatorname{tr}\!\left[
		\big(\mathbf G_m(\mathbf u)
		-\xi_m(\mathbf u)\mathbf I_L\big)\mathbf P_c
		\right],
		\label{51a}\\
		A_m^p(\mathbf u)
		&=
		\operatorname{tr}\!\left[
		\big(\mathbf G_m(\mathbf u)
		-\xi_m(\mathbf u)\mathbf I_L\big)\mathbf P_m
		\right].
		\label{51b}
	\end{align}
\end{subequations}
Applying \textbf{Lemma~1} to these functions yields
\begin{align}
	\widetilde A_m^{s,(i)}(\mathbf u)
	&=
	A_m^s(\mathbf u^{(i)})
	+\nabla A_m^s(\mathbf u^{(i)})^T
	\Delta\mathbf u
	\nonumber\\
	&\quad
	-\frac{\eta_m^{s,(i)}}{2}
	\|\Delta\mathbf u\|_2^2,
	\quad s\in\{c,p\},
	\label{52}
\end{align}
where $\eta_m^{s,(i)}\geq0$ is chosen such that $\nabla^2 A_m^s(\mathbf u) \succeq-\eta_m^{s,(i)}\mathbf I_2$ for every $\mathbf u\in\mathcal C_l^{(i)}$. Since $\widetilde A_m^{s,(i)}(\mathbf u)\leq A_m^s(\mathbf u)$, imposing $\widetilde A_m^{s,(i)}(\mathbf u)\geq0$, $\forall m\in\mathcal M$ and $s\in\{c,p\}$, ensures \eqref{20a} and \eqref{20b}. Moreover, $\widetilde A_m^{s,(i)}$ is concave because its Hessian is $-\eta_m^{s,(i)}\mathbf I_2\preceq\mathbf0$. Hence, these inequalities define convex feasible sets.

When antenna $l$ is updated, the position $\widehat{\boldsymbol t}_j$ of each antenna $j\neq l$
remains fixed. Since $D\geq0$, the separation constraint $\|\mathbf u-\widehat{\boldsymbol t}_j\|_2
\geq D$ is equivalently expressed as $\|\mathbf u-\widehat{\boldsymbol t}_j\|_2^2 \geq D^2$. Since the squared distance is convex, its first-order approximation provides the global lower bound
\begin{align}
	\|\mathbf u-\widehat{\boldsymbol t}_j\|_2^2
	&\geq
	\|\mathbf u^{(i)}-\widehat{\boldsymbol t}_j\|_2^2
	\nonumber\\
	&\quad
	+2(\mathbf u^{(i)}-\widehat{\boldsymbol t}_j)^T
	\Delta\mathbf u.
	\label{53}
\end{align}
Accordingly, imposing the affine constraints
\begin{align}
	&\|\mathbf u^{(i)}-\widehat{\boldsymbol t}_j\|_2^2
	+2(\mathbf u^{(i)}-\widehat{\boldsymbol t}_j)^T
	\Delta\mathbf u
	\geq D^2,
	\nonumber\\
	&\qquad\forall j\in\mathcal L\setminus\{l\},
	\label{54}
\end{align}
ensures the required separation between antenna $l$ and every other antenna. The bound in \eqref{53}
is exact at $\mathbf u=\mathbf u^{(i)}$.

Finally, using \eqref{49a}, \eqref{49b}, \eqref{52}, and \eqref{54}, the update of antenna $l$ and the common-rate allocation is obtained by solving
\begin{subequations}\label{P9}
	\begin{align}
		\text{\textbf{(P9)}}\quad
		&\max_{\mathbf u,\{\widehat r_m^c\}}
		\sum_{m=1}^{M}
		\left(
		\widehat r_m^c+
		\widetilde R_m^{p,(i)}(\mathbf u)
		\right),
		\label{55a}\\
		\mathrm{s.t.}\quad
		&\widehat r_m^c+
		\widetilde R_m^{p,(i)}(\mathbf u)
		\geq R_{\min},
		\quad\forall m\in\mathcal M,
		\label{55b}\\
		&\sum_{k=1}^{M}\widehat r_k^c
		\leq\widetilde R_m^{c,(i)}(\mathbf u),
		\quad\forall m\in\mathcal M,
		\label{55c}\\
		&\widetilde A_m^{s,(i)}(\mathbf u)\geq0,
		\quad\forall m\in\mathcal M,\quad s\in\{c,p\},
		\label{55d}\\
		&\eqref{54},
		\label{55e}\\
		&\mathbf u\in\mathcal C_l^{(i)},
		\label{55f}\\
		&\widehat r_m^c\geq0,
		\quad\forall m\in\mathcal M.
		\label{55g}
	\end{align}
\end{subequations}
Problem \textbf{(P9)} is convex because it maximizes a concave objective over a convex feasible set.
The rate and desired-signal lower bounds, together with \eqref{54}, ensure that every feasible solution of \textbf{(P9)} is feasible for \textbf{(P8)} with the other antenna positions fixed.
Furthermore, if the current design is feasible for \textbf{(P8)}, its position $\mathbf u^{(i)}$
and associated common-rate allocation are feasible for \textbf{(P9)}. This follows because
$\mathbf u^{(i)}\in\mathcal C_l^{(i)}$ and all approximations match their corresponding functions at $\mathbf u^{(i)}$.

	\begin{algorithm}[t]
	\caption{Proposed Robust Optimization Algorithm}
	\label{alg:proposed}
	 \begin{algorithmic}[1]
		\State Initialize a feasible design,
		auxiliary variables, convergence tolerances,
		and iteration limits.
		\Repeat
		\State \textbf{Stage 1: Precoding update}
		\Repeat
		\State Solve the rank-relaxed
		\textbf{(P4)}.
		\State Apply rank handling and
		retain the best feasible design
		under \textbf{(P3)}.
		\State Update the current values of
		$\gamma_m^p,\zeta_m^p,
		\gamma_m^c,\beta_m^c$
		consistently with the retained design.
		\Until{convergence or the iteration limit}
		
		\State \textbf{Stage 2: RIS update}
		\Repeat
		\State Solve the rank-relaxed
		\textbf{(P7)}.
		\State Apply rank handling and
		retain the best feasible design
		under \textbf{(P5)}.
		\State Update $\boldsymbol{\Phi}$,
		$\widehat{\mathbf V}^{(i)}
		=\mathbf v\mathbf v^H$,
		$\mathbf G_m$, and $\xi_m$.
		\Until{convergence or the iteration limit}
		
		\State \textbf{Stage 3: MAs position update}
		\Repeat
		\For{$l=1,\ldots,L$}
		\State Construct and solve
		\textbf{(P9)} with the other
		antenna positions fixed.
		\State Update
		$\widehat{\boldsymbol t}_l$,
		$\mathbf G_m$, and $\xi_m$.
		\EndFor
		\Until{convergence or the iteration limit}
		\Until{sum-rate convergence or
			the outer iteration limit}
		\State \Return $\mathbf p_c$,
		$\{\mathbf p_m\}$, $\boldsymbol{\Phi}$,
		$\widehat{\boldsymbol t}$, and
		$\{\widehat r_m^c\}$.
	\end{algorithmic}
\end{algorithm}

\subsection{Analysis of the Proposed Robust Algorithm}   
\noindent\textit{1) Computational Complexity:} The computational cost of Algorithm~1 is dominated by
the repeated solutions of the rank-relaxed problems \textbf{(P4)} and \textbf{(P7)}, and the convex
single-antenna position problem \textbf{(P9)}. We estimate these costs using a conservative dense
interior-point analysis, with the solver accuracy held fixed. The analysis combines the barrier-based
iteration bound \cite{vandenberghe2004convex,papp2022alfonso} with the cost of dense linear-system factorization \cite{vandenberghe2004convex}.

The precoding problem \textbf{(P4)} contains $M+1$ Hermitian matrices of size $L\times L$ and $6M$
real scalar variables. Its conic formulation has size $\mathcal O((M+1)L^2+M)$. The $M+1$ PSD constraints contribute $\mathcal O((M+1)L)$ to the barrier parameter, while the remaining scalar,
second-order-cone, and exponential-cone constraints contribute $\mathcal O(M)$. The total barrier
parameter is therefore $\mathcal O((M+1)L+M)$ \cite{vandenberghe2004convex,papp2022alfonso}. Combining the interior-point iteration bound with the dense per-iteration cost gives the per-solution complexity $\mathcal C_{\mathrm p} =\mathcal O\!\left(\sqrt{(M+1)L+M}\,[(M+1)L^2+M]^3\right)$.

Similarly, the RIS problem \textbf{(P7)} contains one Hermitian matrix of size $N\times N$ and $2M$ real scalar variables. Its conic formulation has size $\mathcal O(N^2+M)$. The PSD constraint contributes order $N$ to the barrier parameter, while the remaining cone constraints contribute $\mathcal O(M)$, giving a total barrier parameter of $\mathcal O(N+M)$. Combining the interior-point
iteration bound with the dense per-iteration cost gives the per-solution complexity $\mathcal C_{\mathrm r}=\mathcal O\!\left(\sqrt{N+M}\,(N^2+M)^3\right).$

The single-antenna position problem \textbf{(P9)} optimizes two position coordinates and $M$ common-rate allocations. It includes $\mathcal O(M)$ user-related constraints and $L-1$ antenna-separation constraints. Assuming a fixed-size conic representation of the search region, its conic formulation has size $\mathcal O(M+L)$. The user-related constraints contribute $\mathcal O(M)$ to the barrier parameter, while the affine antenna-separation constraints contribute $\mathcal O(L)$ and the search-region constraints contribute $\mathcal O(1)$. The total barrier parameter is therefore
$\mathcal O(M+L)$. Combining the interior-point iteration bound of order $\sqrt{M+L}$ with
the conservative dense per-iteration cost $\mathcal O((M+L)^3)$ gives the per-solution complexity $\mathcal C_{\mathrm t}=\mathcal O\!\left((M+L)^{3.5}\right).$

Let $\varepsilon_1$ and $\varepsilon_2$ denote upper bounds on the numbers of precoding and RIS
iterations per outer iteration, respectively, and let $\varepsilon_3$ denote an upper bound on the number of complete antenna sweeps. Since a complete sweep solves \textbf{(P9)} for all $L$ antennas, the overall convex-solver complexity is estimated as $\mathcal O\!\left(\hat{\chi}_{\max}\left[ \varepsilon_1\mathcal C_{\mathrm p}+\varepsilon_2\mathcal C_{\mathrm r}+\varepsilon_3L\mathcal C_{\mathrm t}\right]\right)$, where $\hat{\chi}_{\max}$ is the maximum number of outer iterations.

\noindent\textit{2) Convergence Analysis:}
Let $\mathbf z$ collect the precoding vectors, RIS reflection matrix, MA positions, and common-rate allocations, and define the conservative sum-rate as $\Xi(\mathbf z)=\sum_{m=1}^{M}(\widehat r_m^c+\overline R_m^p)$. At outer iteration $r$, let $\mathbf z_0^{(r)}$ denote the design before the three updates and $\mathbf z_j^{(r)}$ denote the design after Stage $j$, $j\in\{1,2,3\}$. Thus, $\mathbf z_0^{(r+1)}=\mathbf z_3^{(r)}$.

In Stage~1, the rank-relaxed problem \textbf{(P4)} is solved, and precoding vectors are obtained
through eigenvalue decomposition for rank-one solutions or Gaussian randomization otherwise. The candidates and their associated common-rate allocations are checked against \textbf{(P3)}. Since the previous feasible design is included among the candidates, selecting the design with the largest conservative sum-rate ensures
\begin{equation}
	\Xi(\mathbf z_1^{(r)})
	\geq
	\Xi(\mathbf z_0^{(r)}).
	\label{56}
\end{equation}
This acceptance rule preserves monotonicity even when rank-one recovery does not produce an improving feasible candidate.

Similarly, Stage~2 solves the rank-relaxed problem \textbf{(P7)} and checks the resulting RIS candidates against \textbf{(P5)}. Retaining the best feasible design, including the previous design, gives
\begin{equation}
	\Xi(\mathbf z_2^{(r)})
	\geq
	\Xi(\mathbf z_1^{(r)}).
	\label{57}
\end{equation}
The objective is evaluated using the recovered unit-modulus RIS vector, with $\mathbf G_m$
and $\xi_m$ recomputed accordingly.

In Stage~3, the antennas are updated sequentially by solving \textbf{(P9)}. With the other design
variables fixed, write the conservative sum-rate as $\Xi(\mathbf u,\widehat{\mathbf r}^c) =\sum_{m=1}^{M} (\widehat r_m^c+\overline R_m^p(\mathbf u))$ and define its surrogate as
$\widetilde\Xi^{(i)}(\mathbf u,\widehat{\mathbf r}^c)=\sum_{m=1}^{M} (\widehat r_m^c+\widetilde R_m^{p,(i)}(\mathbf u))$, where $\widehat{\mathbf r}^c$ collects the common-rate allocations. The surrogate lower-bounds the conservative sum-rate and matches it at $\mathbf u^{(i)}$ for the same common-rate allocation. Moreover, the current position and its associated common-rate allocation remain feasible for \textbf{(P9)}. Let $\mathbf u_{\mathrm{new}}$ and $\widehat{\mathbf r}_{\mathrm{new}}^c$ denote the solution of \textbf{(P9)}, and let $\widehat{\mathbf r}_{\mathrm{old}}^c$ denote the allocation before the update. Solving
\textbf{(P9)} to optimality therefore yields
\[
\begin{aligned}
	\Xi(\mathbf u_{\mathrm{new}},
	\widehat{\mathbf r}_{\mathrm{new}}^c)
	&\geq
	\widetilde\Xi^{(i)}
	(\mathbf u_{\mathrm{new}},
	\widehat{\mathbf r}_{\mathrm{new}}^c)\\
	&\geq
	\widetilde\Xi^{(i)}
	(\mathbf u^{(i)},
	\widehat{\mathbf r}_{\mathrm{old}}^c)\\
	&=
	\Xi(\mathbf u^{(i)},
	\widehat{\mathbf r}_{\mathrm{old}}^c).
\end{aligned}
\]
The surrogate constraints in \textbf{(P9)} are conservative restrictions of the corresponding constraints in \textbf{(P8)}. Thus, the solution of \textbf{(P9)} satisfies the constraints of \textbf{(P8)} with the other antenna positions fixed. Together with the inequality above, this shows that each single-antenna update preserves feasibility and does not decrease the conservative sum-rate. The complete position-update stage consequently satisfies
\begin{equation}
	\Xi(\mathbf z_3^{(r)})
	\geq
	\Xi(\mathbf z_2^{(r)}).
	\label{58}
\end{equation}

Combining \eqref{56}--\eqref{58} gives
\begin{align}
	\Xi(\mathbf z_0^{(r+1)})
	&=\Xi(\mathbf z_3^{(r)})
	\geq\Xi(\mathbf z_2^{(r)})
	\nonumber\\
	&\geq\Xi(\mathbf z_1^{(r)})
	\geq\Xi(\mathbf z_0^{(r)}).
	\label{59}
\end{align}
Thus, the conservative sum-rate is monotonically non-decreasing. Under the finite transmit-power
budget, bounded channel gains over the compact position region, and strictly positive noise powers, the objective is upper bounded. Consequently, the sequence of conservative sum-rate values generated by Algorithm~1 converges.

\noindent\textit{3) Interaction Among Optimization Stages and Update Order:}
The active beamforming, RIS reflection coefficients, and MAs positions are jointly optimized to improve the conservative sum-rate while satisfying the system constraints. With the RIS coefficients and MA positions fixed, the active beamforming update adjusts the precoding vectors and common-rate allocation to improve the conservative sum-rate. The RIS update then modifies the reflected channels for the current transmit design, while the MA-position update adjusts the spatial channels through antenna movement. Both updates also change the uncertainty radii through $\xi_m=\rho\|\mathbf G_m\|_2$. The updated channels and uncertainty radii are subsequently used in the next active beamforming update. The common-rate allocation is jointly optimized in each stage to satisfy the communication requirements.

The adopted order first updates the active beamforming, then the RIS coefficients, and finally the MA positions. The monotonic convergence argument does not require this particular order. It remains valid for alternative orders provided that each stage preserves feasibility and does not decrease the same conservative sum-rate. However, changing the update order changes the intermediate designs available to subsequent stages. Since the joint problem is non-convex, this may affect both the number of outer iterations required for convergence and the final achieved sum-rate. The numerical evaluations consistently follow the adopted update order.

\noindent\textit{4) Scalability:} For fixed $M$ and $L$, increasing the number of RIS elements $N$ enlarges the $N\times N$ lifted reflection matrix in \textbf{(P7)}. Under the adopted dense interior-point model, the computational cost per solve scales as $\mathcal O(N^{6.5})$, while storage for the lifted matrix scales as $\mathcal O(N^2)$. For fixed $N$ and $L$, increasing the number of users $M$ adds one private-stream covariance matrix per user in \textbf{(P4)} and increases the number of user-specific variables and constraints across all three subproblems. In this regime, the per-solve complexity estimates for \textbf{(P4)}, \textbf{(P7)}, and \textbf{(P9)} each scale as $\mathcal O(M^{3.5})$. Thus, for fixed $M$ and $L$, increasing $N$ raises the estimated per-solve complexity of the lifted RIS subproblem \textbf{(P7)}, while for fixed $N$ and $L$, increasing $M$ raises the estimated per-solve complexity of all three subproblems.

\section{Simulation Results and Performance Analysis}
\begin{figure}[!t]
	\centering
	\includegraphics [width=0.46\textwidth]{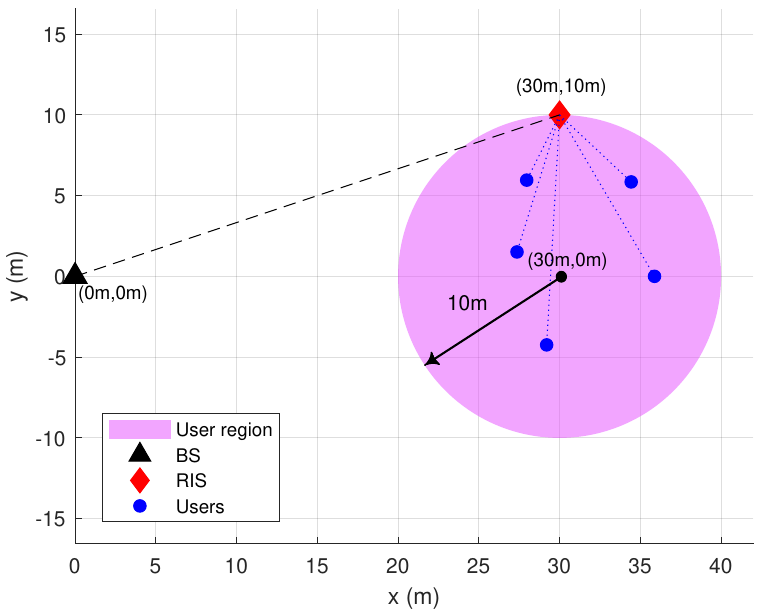}
	\caption{Simulation environment.}
	\label{f2}
\end{figure}

In this work, the performance of the considered system is evaluated through 
numerical simulations. As illustrated in Fig.~\ref{f2}, the BS and RIS are 
located at coordinates $(0~\text{m}, 0~\text{m})$ and $(30~\text{m}, 10~\text{m})$, respectively. Moreover, the users are randomly distributed within a circular region of radius $10~\text{m}$, centered at $(30~\text{m}, 0~\text{m})$. Further, the planar far-field geometric channel model is adopted with $L_t=L_r$, where the elevation and azimuth angles of all propagation paths, $\phi_i^{e}$, $\varphi_j^{e}$, $\phi_i^{a}$, and $\varphi_j^{a}$, are uniformly distributed over $[0,\pi]$. Unless otherwise specified, all simulation results are averaged over $10^3$ independent channel realizations. It is assumed that line-of-sight (LoS) links exist for both the BS--RIS and RIS--user channels. Accordingly, the diagonal elements of the BS--RIS path response matrix $\boldsymbol{\Lambda}$ are modeled as $\boldsymbol{\Lambda}[1,1]\sim\mathcal{CN}\!\left(0,\frac{\widehat{\varsigma}}{\widehat{\varsigma}+1}\mathcal{P}_0\!\left(\frac{d}{d_0}\right)^{-\widehat{\nu_1}}\right)$ and $\boldsymbol{\Lambda}[i,i]\sim\mathcal{CN}\!\left(0,\frac{1}{\widehat{\varsigma}+1}\mathcal{P}_0\!\left(\frac{d}{d_0}\right)^{-\widehat{\nu_1}}/(L-1)\right)$ for $i= 2,3,...,L$, where $\widehat{\varsigma}$ and $\widehat{\nu_1}$ denote the Rician factor and path-loss exponent, respectively, and $\mathcal{P}_0$ represents the average channel power gain at a reference distance $d_0=1~\text{m}$. Moreover, the RIS--user channels $\mathbf{h}_{r,m}\in\mathbb{C}^{N\times 1}$ are independent of the MA positions and are modeled using Rician fading to account for both LoS and non-line-of-sight (NLoS) components, where path-loss exponent for RIS--user link is denoted as $\widehat{\nu_2}$. Unless otherwise specified, the main simulation parameters are configured as follows: $\widehat{\nu_1}=2$, $\widehat{\nu_2}=2.5$, $J_{\mathrm p}=J_{\mathrm r}=10^3$, $\mathcal{P}_0=-30~\text{dB}$, $\mathcal{C}_t=\left[-\frac{A}{2}, \frac{A}{2}\right] \times\left[-\frac{A}{2}, \frac{A}{2}\right]$, $A=3\lambda$ with $\lambda=0.1~\text{m}$ \cite{xiao2024multiuser}, $\widehat{{\sigma}}_m^2 = -114~\text{dBm}$, $R_{\min} = 1~ \text{bps/Hz}$, $D=\lambda/2$, $P_t= 15~\text{Watts}$, $\widehat{\varsigma}=2$, $M=4$, and $\xi_m=\rho\|\mathbf G_m\|_2$, where $\rho \in [0, 1)$ \cite{zheng2023zero}.

The proposed scheme, denoted by MA-RIS throughout the numerical results, is compared with the following four benchmark schemes. At each comparison point, all schemes use the same underlying channel realizations, transmit power budget $P_t$, number of RIS elements $N$, number of BS antennas $L$, and CSI uncertainty model with the same relative uncertainty level $\rho$.  All schemes that optimize antenna positions use the same movement region $\mathcal C_t$ and minimum antenna separation $D$.

\begin{itemize}
	\item \textbf{FPA-RIS:} The BS antenna positions are fixed, while the RSMA precoders, common-rate allocation, and RIS reflection coefficients are optimized. This benchmark evaluates the benefit of antenna-position optimization.
	
	\item \textbf{SDMA-RIS:} The BS transmits only private streams. The transmit precoders, RIS reflection coefficients, and MA positions are jointly optimized. This benchmark evaluates the benefit of rate splitting.
	
   \item \textbf{PSO-RIS:} This benchmark employs particle swarm optimization (PSO) \cite{ding2024movable}to perform a joint search over all design variables. Each particle represents a candidate solution, and its position is updated using the best solution it has encountered and the best solution identified by the swarm. The candidates are evaluated under the same conservative sum-rate objective and design constraints as the proposed scheme.
	
   \item \textbf{CWS-RIS:} This benchmark employs coordinate-wise search (CWS), adapting the alternating position selection strategy in \cite{xiao2024multiuser} to the full set of design variables. At each step, it tests changes to one coordinate while keeping the others fixed and accepts a change only if it preserves feasibility and improves the conservative sum-rate. The search step sizes are gradually reduced until the stopping criterion is satisfied.
\end{itemize}

\begin{figure}[!t]
	\centering
	\includegraphics [width=0.46\textwidth]{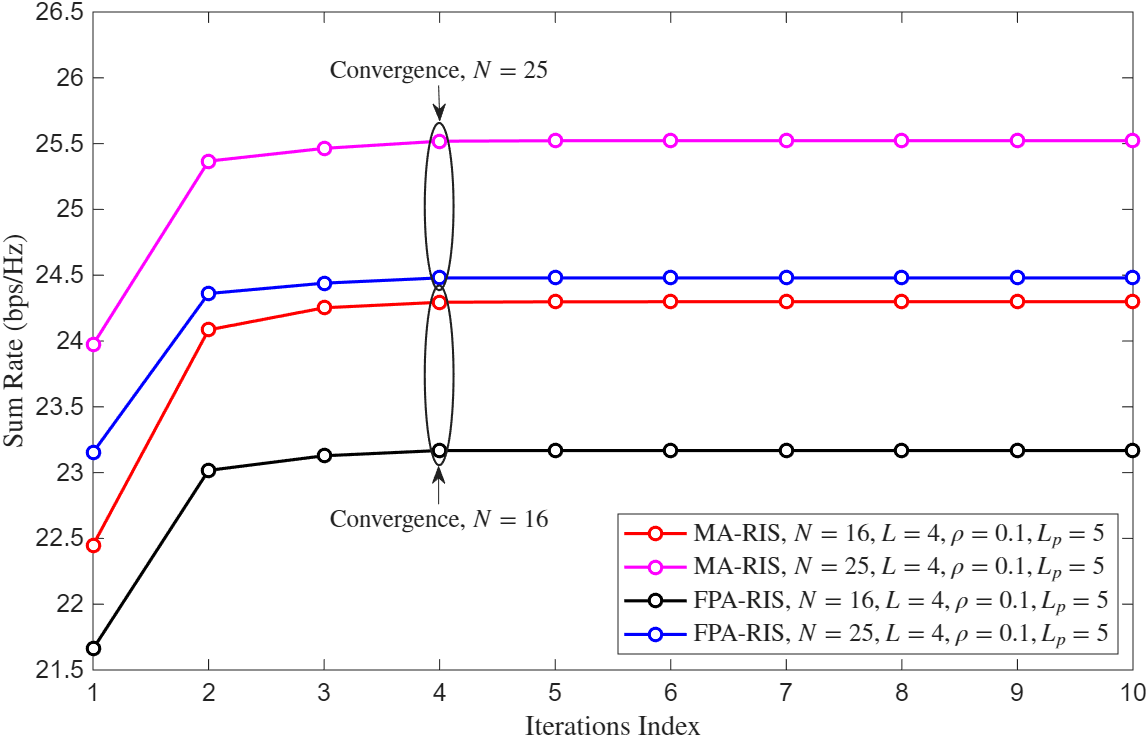}
	\caption{System convergence for different number of RIS elements.}
	\label{f3}
\end{figure} 

  \begin{figure}[!t]
	\centering
	\includegraphics [width=0.46\textwidth]{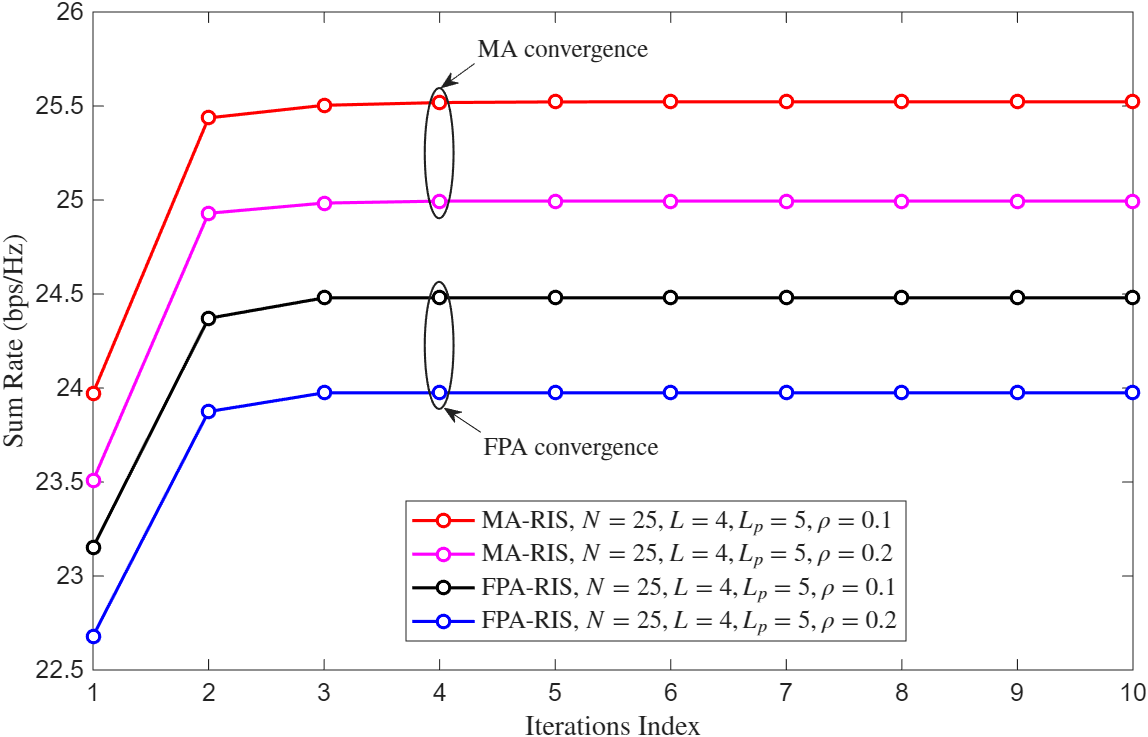}
	\caption{System convergence for different levels of channel uncertainty.}
	\label{f4}
\end{figure}

  \begin{figure}[!t]
	\centering
	\includegraphics [width=0.46\textwidth]{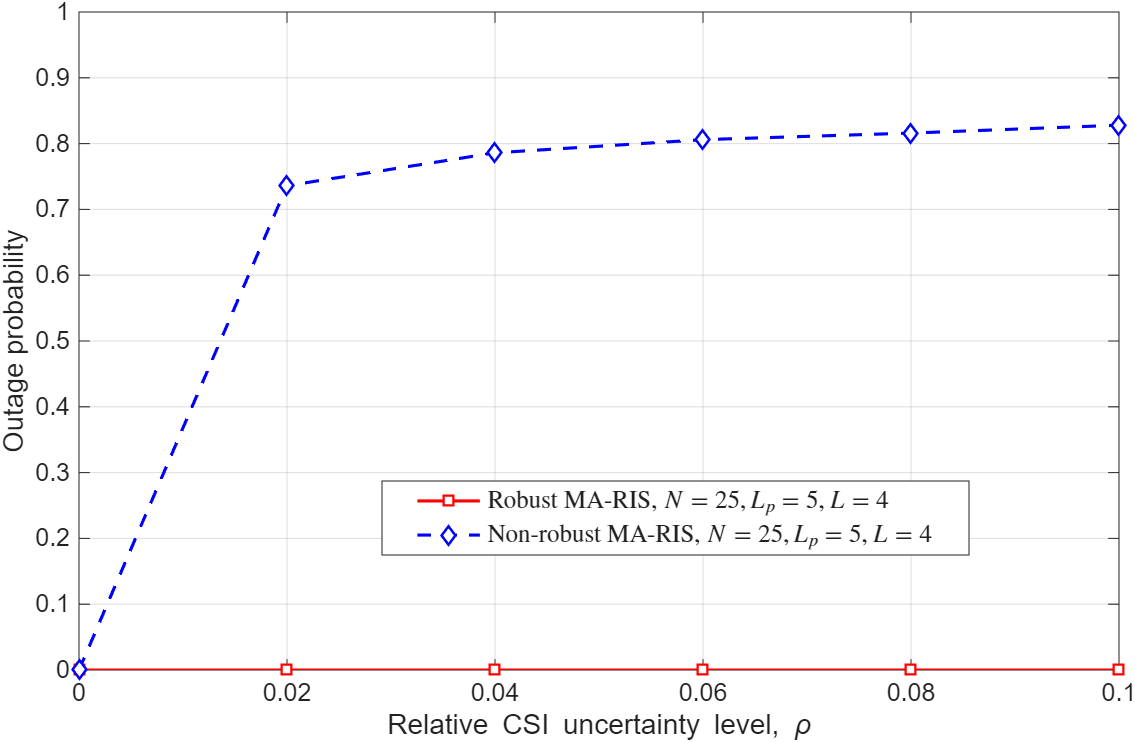}
	\caption{System outage probability versus relative CSI uncertainty $\rho$ for robust and non-robust MA-RIS.}
	\label{f5}
\end{figure}

  \begin{figure}[!t]
	\centering
	\includegraphics [width=0.46\textwidth]{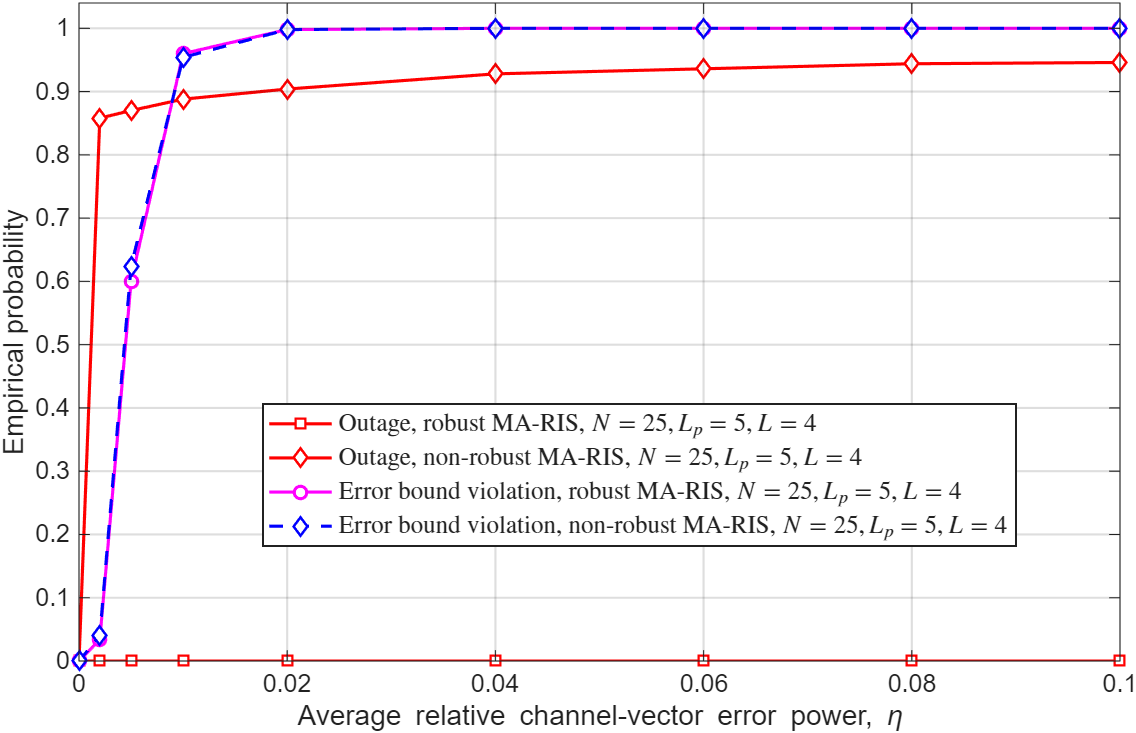}
	\caption{Empirical outage and uncertainty-bound violation probabilities under Gaussian channel-estimation errors.}
	\label{f6}
\end{figure}

\begin{figure}[t]
	\centering
	\includegraphics [width=0.46\textwidth]{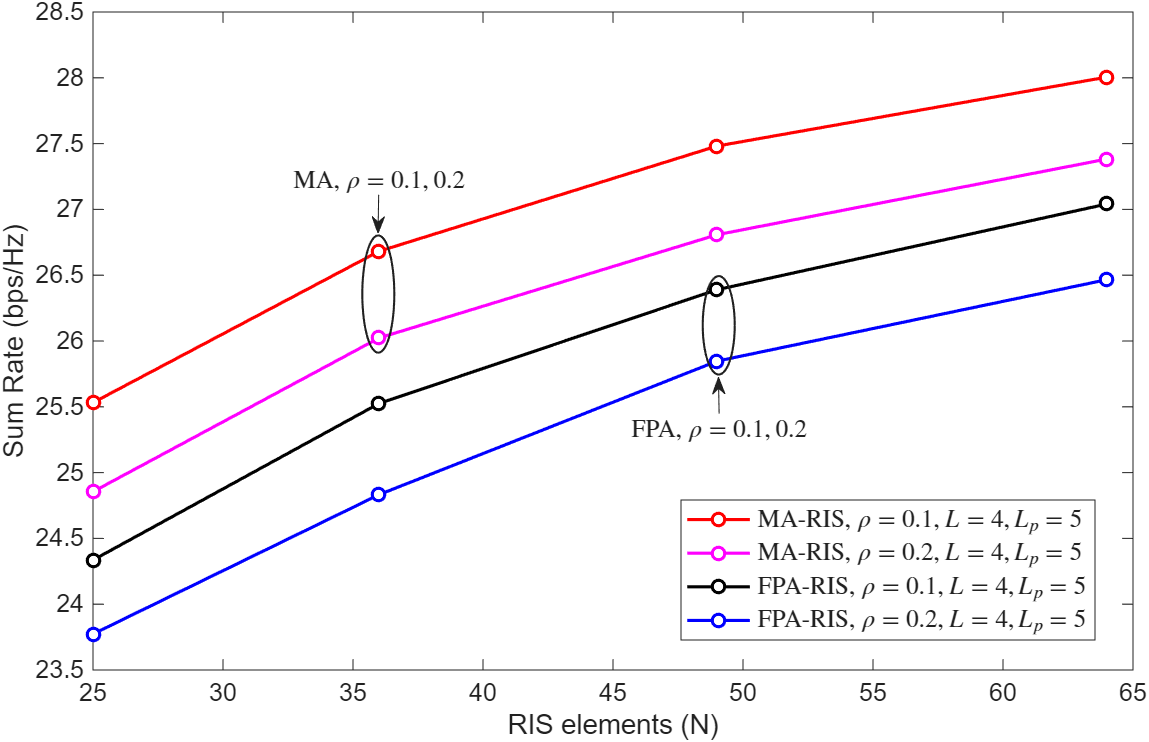}
	\caption{Sum-rate performance across increasing values of $N$.}
	\label{f7}
\end{figure}

\begin{figure}[t]
\centering
\includegraphics [width=0.46\textwidth]{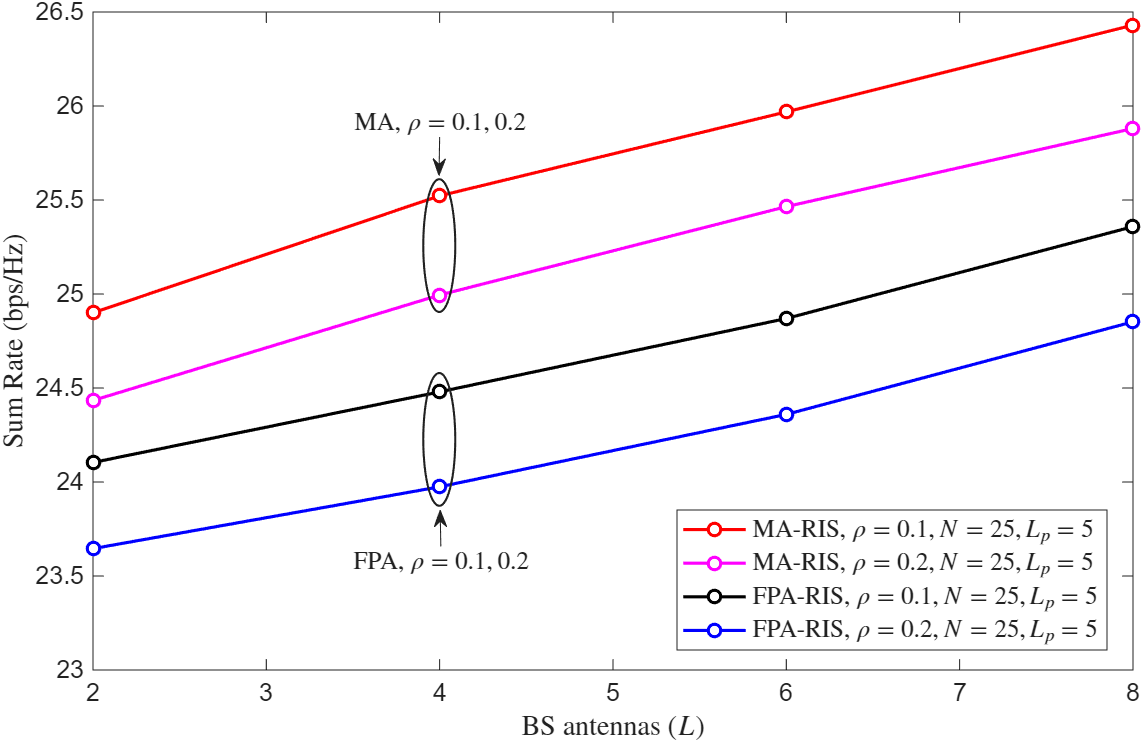}
\caption{Sum-rate performance for different number of MAs $L$.}
\label{f8}
\end{figure}

\begin{figure}[t]
	\centering
	\includegraphics [width=0.46\textwidth]{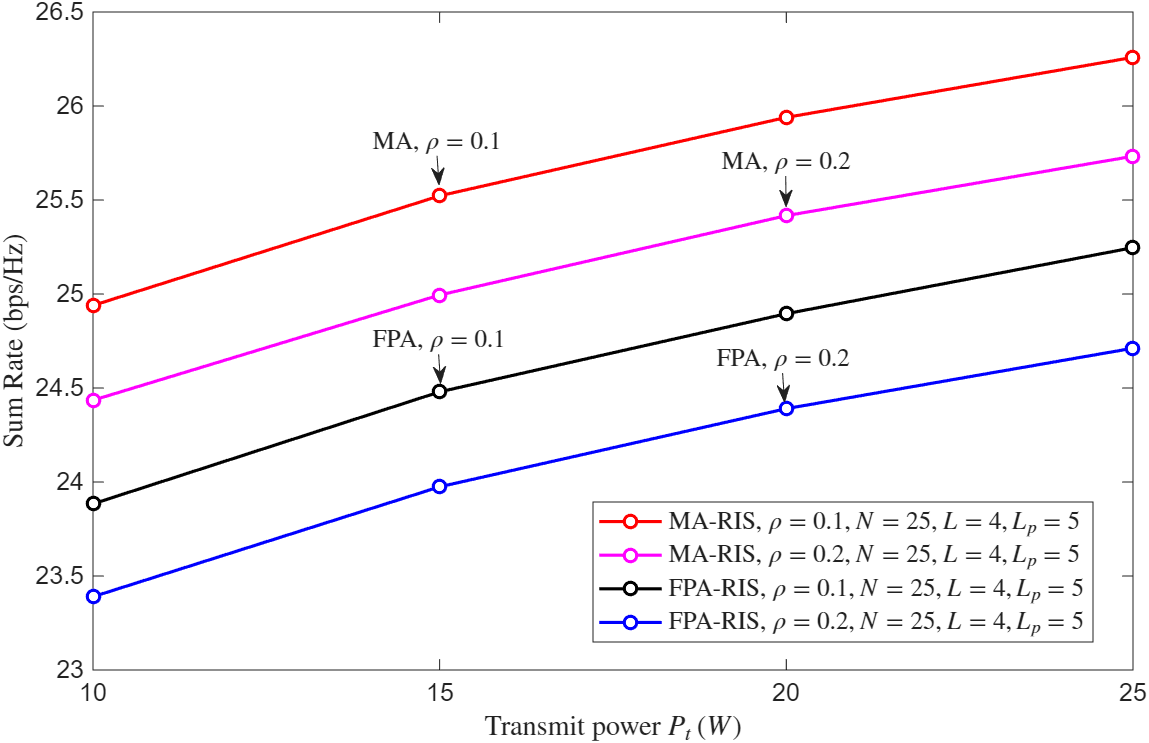}
	\caption{Performance analysis for increasing power-budget $P_t$.}
	\label{f9}
\end{figure}

Figs.~\ref{f3} and~\ref{f4} compare the convergence behavior of the proposed MA-RIS scheme and the FPA-RIS benchmark under different system parameter settings. Fig.~\ref{f3} shows how the sum-rate evolves over successive iterations for different numbers of RIS elements $N$, while Fig.~\ref{f4} shows the corresponding results for different values of the channel uncertainty parameter $\rho$. For every parameter setting shown, the proposed MA-RIS scheme converges to a higher sum-rate than the FPA-RIS benchmark.

Fig.~\ref{f5} compares the empirical outage probabilities of robust and non-robust MA-RIS as a function of the relative CSI uncertainty level $\rho$. For each estimated channel realization, the robust design is optimized at each tested value of $\rho$, whereas the non-robust design is optimized once using the estimated CSI with $\rho=0$. Both designs are evaluated at the same test value of $\rho$ using matched random error draws. The draws are scaled according to each design's estimated effective channel so that the resulting perturbations satisfy the prescribed uncertainty bound. We estimate the system outage probability as the fraction of evaluated trials in which at least one user fails to decode the common stream at the total allocated common rate or fails to meet its minimum rate requirement. The numerical simulations yield an empirical outage probability of zero for robust MA-RIS over the tested range, whereas the outage probability of non-robust MA-RIS reaches approximately $0.83$ at $\rho=0.10$. These results show that accounting for CSI uncertainty during optimization reduces rate constraint violations under the tested channel perturbations.

Fig.~\ref{f6} evaluates the robustness of MA-RIS under unbounded Gaussian channel-estimation errors. The robust and non-robust designs are optimized with $\rho=0.10$ and $\rho=0$, respectively, and subsequently held fixed during evaluation. For each user, the test error is generated as $\Delta\mathbf g_m=\sqrt{\eta/L}\|\mathbf g_m\|_2\mathbf w_m$, where $\mathbf w_m\sim\mathcal{CN}(\mathbf0,\mathbf I_L)$ and $\eta$ denotes the average relative channel-vector error power. Both designs use matched Gaussian draws scaled by their respective estimated effective channels. All generated errors are retained, including those whose induced matrix perturbations exceed the reference bound. Alongside the system outage probability defined for Fig.~\ref{f5}, we report the probability that at least one user satisfies $\|\Delta\mathbf G_m\|_2>0.10\|\mathbf G_m\|_2$, using the same reference relative uncertainty level for both designs. As $\eta$ increases, the empirical bound-violation probability approaches one for both designs. Nevertheless, robust MA-RIS maintains a substantially lower outage probability than non-robust MA-RIS. At $\eta=0.10$, all 500 evaluated trials contain an uncertainty-bound violation for at least one user, while the empirical outage probabilities are zero and $0.946$ for robust and non-robust MA-RIS, respectively. These results demonstrate the robustness of the proposed design under the tested Gaussian perturbations, including those outside the prescribed uncertainty set, although its theoretical rate guarantees apply only within that set.

Fig.~\ref{7} exhibits the impact of the increasing number of RIS elements $N$ on the system sum-rate under different values of $\rho$. It can be seen that the sum-rate performance improves with an increase in $N$. This is due to the increased effective channel gain and the additional spatial degrees of freedom provided by the RIS, which improve the effectiveness of passive beamforming. Also, it is clear that the proposed scheme always outperforms the FPA-RIS benchmark in terms of sum-rate performance for all values of $N$ and $\rho$. On the other hand, the sum-rate performance decreases with an increase in $\rho$. This can be attributed to the negative effect of high channel uncertainty on channel estimation accuracy.

Fig.~\ref{f8} shows the effect of the number of movable antennas $L$ on the sum-rate performance of the system for different values of the channel uncertainty parameter $\rho$. It is observed that the increase in $L$ results in a guaranteed improvement in the sum-rate performance. This is because the movable antennas introduce an extra degree of spatial freedom that helps in more optimal positioning of the antennas and, hence, a better beamforming gain at the BS. This results in a better focusing of the signal power towards the desired receivers. However, the sum-rate performance deteriorates with an increase in $\rho$ due to the negative effect of the increased channel uncertainty on the SINR values.

\begin{figure}[t]
	\centering
	\includegraphics [width=0.46\textwidth]{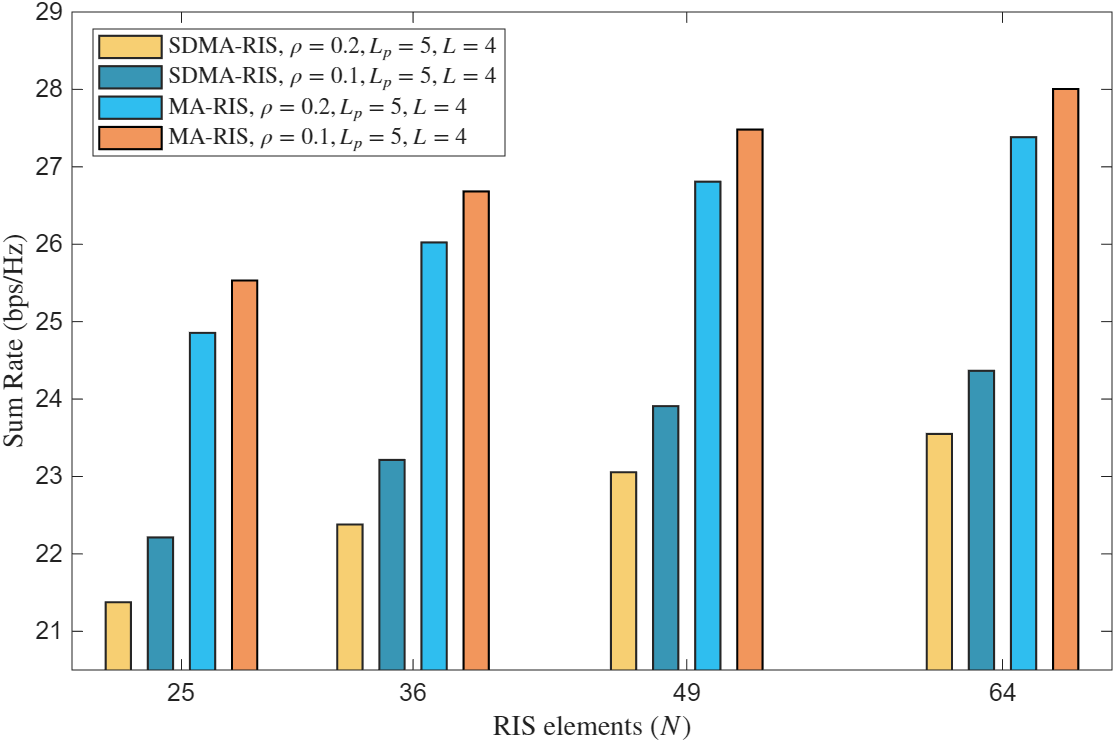}
	\caption{Performance comparison of proposed scheme and its SDMA benchmark.}
	\label{f10}
\end{figure}

Fig.~\ref{f9} depicts the variation of the achievable system sum-rate with respect to the transmit power budget $P_t$ under different levels of channel uncertainty $\rho$. As $P_t$ increases, a steady improvement in sum-rate performance is observed, which can be attributed to the enhanced received signal strength and the resulting increase in the SINR at the users. Moreover, the proposed scheme consistently achieves higher sum-rate performance than the FPA-RIS benchmark for all considered values of $P_t$ and $\rho$, demonstrating its superior robustness and power utilization efficiency under imperfect CSI conditions. Moreover, Fig.~\ref{f10} presents a comparative evaluation of the proposed scheme and its SDMA-based benchmark, referred to as SDMA-RIS, under varying numbers of RIS elements $N$ and different channel uncertainty levels $\rho$. The results indicate that the proposed approach consistently achieves higher sum-rate performance across all considered values of $N$ and $\rho$. This gain stems from the flexible interference management provided by RSMA, which allows users to decode part of the multiuser interference through the common stream while treating the remaining interference as noise. Jointly optimizing the common and private precoders and common-rate allocation provides additional flexibility to manage interference under CSI uncertainty.

\begin{figure}[t]
	\centering
	\includegraphics [width=0.46\textwidth]{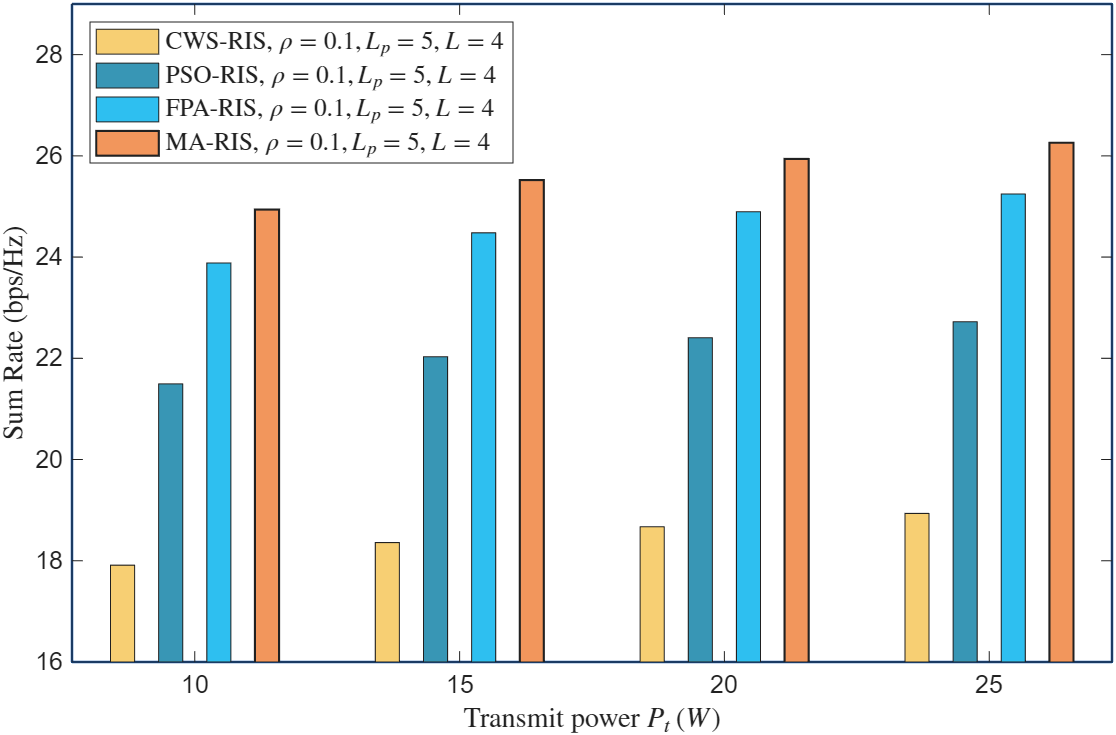}
	\caption{Performance Comparison against benchmarks for increasing values of $P_t$.}
	\label{f11}
\end{figure} 

Fig.~\ref{f11} compares the conservative sum-rate of MA-RIS with the FPA-RIS, PSO-RIS, and CWS-RIS benchmarks as the transmit-power budget $P_t$ increases. All schemes achieve higher sum-rates with increasing $P_t$, owing to the additional power available for transmission. MA-RIS consistently provides the highest sum-rate, followed by FPA-RIS, PSO-RIS, and CWS-RIS. The gain over FPA-RIS demonstrates the benefit of adapting the MAs positions jointly with the transmit precoders and RIS reflection coefficients. Its advantage over PSO-RIS and CWS-RIS, which optimize the same design variables, illustrates the effectiveness of the proposed algorithm under the considered simulation settings. Moreover, FPA-RIS achieves higher sum-rates than PSO-RIS and CWS-RIS benchmarks despite using fixed antenna positions, highlighting the importance of the optimization method in exploiting the available design variables under the considered simulation settings.

\section{Conclusion}
This paper proposed a robust joint resource allocation framework for an RIS-empowered MAs-enabled multi-user RSMA downlink system under channel estimation errors. By jointly exploiting the spatial adaptability of MAs, the propagation reconfigurability of RIS, and the interference management capability of RSMA, the proposed framework enhances the system sum-rate while satisfying QoS requirements in the absence of direct communication links. In particular, changes in the MAs positions and RIS reflection coefficients affect both the estimated effective channels and the corresponding uncertainty radii under the adopted uncertainty model, introducing a coupled robust optimization structure. To address this challenge, an iterative optimization framework was developed to jointly optimize active beamforming, RIS reflection control, common-rate allocation, and MAs positions while enforcing the rate requirements within the adopted uncertainty sets. Numerical results demonstrated that the proposed framework achieves significant sum-rate gains and enhanced robustness compared with benchmark schemes, while also exhibiting fast and stable convergence behavior. Additional outage evaluations under bounded channel-estimation errors and unbounded Gaussian errors demonstrated improved robustness in satisfying user-rate requirements. The proposed framework provides useful insights into robust and spatially adaptive transmission design under imperfect CSI. Future work may extend the framework to dynamic channel conditions while explicitly accounting for antenna movement latency and repositioning energy consumption.

\section*{\textbf{Appendix A: Proof of Theorem 1}}
Consider two complex matrices $\widehat{\mathbf\Lambda}$ and $\widehat{\mathbf\Omega}$ of the same dimensions. Define their real inner product as $\langle\widehat{\mathbf\Lambda}, \widehat{\mathbf\Omega}\rangle =\operatorname{Re}\{\operatorname{tr}(\widehat{\mathbf\Lambda}^{H}
\widehat{\mathbf\Omega})\}$. By the duality between the spectral and nuclear norms,
\begin{equation}
	\max_{\|\widehat{\mathbf\Lambda}\|_2\leq1}
	\langle\widehat{\mathbf\Lambda},
	\widehat{\mathbf\Omega}\rangle
	=
	\|\widehat{\mathbf\Omega}\|_{\mathrm{nuclear}}.
	\label{A-51}
\end{equation}
The corresponding trace inequality is
\begin{equation}
	\left|
	\operatorname{tr}\!\left(
	\widehat{\mathbf\Omega}^{H}
	\widehat{\mathbf\Lambda}\right)
	\right|
	\leq
	\|\widehat{\mathbf\Lambda}\|_2
	\|\widehat{\mathbf\Omega}\|_{\mathrm{nuclear}}.
	\label{A-52}
\end{equation}

Now, let $\widehat{\mathbf\Psi}$ and $\widehat{\mathbf\Xi}$ be Hermitian matrices, with
$\widehat{\mathbf\Psi}\succeq\mathbf0$ and $\|\widehat{\mathbf\Xi}\|_2\leq\xi_m$. Their trace product is real. Applying \eqref{A-52} gives
\begin{align}
	\operatorname{tr}\!\left(
	\widehat{\mathbf\Psi}\widehat{\mathbf\Xi}\right)
	&\leq
	\left|\operatorname{tr}\!\left(
	\widehat{\mathbf\Psi}\widehat{\mathbf\Xi}\right)\right|
	\leq
	\|\widehat{\mathbf\Xi}\|_2
	\|\widehat{\mathbf\Psi}\|_{\mathrm{nuclear}}
	\nonumber\\
	&\leq
	\xi_m\|\widehat{\mathbf\Psi}\|_{\mathrm{nuclear}}.
	\label{A-53}
\end{align}
Consequently,
\begin{equation}
	\max_{\|\widehat{\mathbf\Xi}\|_2\leq\xi_m}
	\operatorname{tr}\!\left(
	\widehat{\mathbf\Psi}\widehat{\mathbf\Xi}\right)
	\leq
	\xi_m\|\widehat{\mathbf\Psi}\|_{\mathrm{nuclear}},
	\label{A-54}
\end{equation}
where the maximization is over Hermitian matrices
$\widehat{\mathbf\Xi}$.

Since $\widehat{\mathbf\Psi}$ is positive semidefinite, its singular values coincide with its nonnegative eigenvalues. Therefore,
\begin{equation}
	\|\widehat{\mathbf\Psi}\|_{\mathrm{nuclear}}
	=
	\sum_i\varsigma_i
	=
	\operatorname{tr}(\widehat{\mathbf\Psi}),
	\label{A-55}
\end{equation}
where $\{\varsigma_i\}$ are the eigenvalues of $\widehat{\mathbf\Psi}$. This equality holds for any positive semidefinite matrix, irrespective of its rank.

Finally, the choice $\widehat{\mathbf\Xi}=\xi_m\mathbf I$ is Hermitian, satisfies $\|\widehat{\mathbf\Xi}\|_2\leq\xi_m$, and attains the upper bound in \eqref{A-54}. Hence,
\begin{equation}
	\max_{\|\widehat{\mathbf\Xi}\|_2\leq\xi_m}
	\operatorname{tr}\!\left(
	\widehat{\mathbf\Psi}\widehat{\mathbf\Xi}\right)
	=
	\xi_m\operatorname{tr}(\widehat{\mathbf\Psi}).
	\label{A-56}
\end{equation}
This completes the proof.

\section*{\textbf{Appendix B: Proof of Lemma 1}}
Let $\Delta\mathbf u=\mathbf u-\mathbf u^{(i)}$. Parameterize the segment from $\mathbf u^{(i)}$ to
$\mathbf u$ by $\mathbf u^{(i)}+\tau\Delta\mathbf u$, where $\tau\in[0,1]$ is a real scalar. The choices $\tau=0$ and $\tau=1$ correspond to $\mathbf u^{(i)}$ and $\mathbf u$, respectively. Since $\mathcal C_l^{(i)}$ is convex, this entire segment lies within $\mathcal C_l^{(i)}$.

Further, define the real-valued scalar function $g(\tau)=f(\mathbf u^{(i)}+\tau\Delta\mathbf u)$.
By the chain rule,
\[
\begin{aligned}
	g'(\tau)
	&=
	\nabla f(\mathbf u^{(i)}+\tau\Delta\mathbf u)^T
	\Delta\mathbf u,\\
	g''(\tau)
	&=
	\Delta\mathbf u^T
	\nabla^2 f(\mathbf u^{(i)}+\tau\Delta\mathbf u)
	\Delta\mathbf u.
\end{aligned}
\]
Consequently, the Hessian bound in \eqref{47} implies
\[
g''(\tau)
\geq
-\chi_f^{(i)}\|\Delta\mathbf u\|_2^2,
\qquad \forall\tau\in[0,1].
\]

Integration by parts gives
\[
\begin{aligned}
	\int_0^1(1-\tau)g''(\tau)\,d\tau
	&=
	\big[(1-\tau)g'(\tau)\big]_0^1
	+\int_0^1g'(\tau)\,d\tau\\
	&=
	-g'(0)+g(1)-g(0).
\end{aligned}
\]
Using $g(0)=f(\mathbf u^{(i)})$, $g(1)=f(\mathbf u)$, and $g'(0)=\nabla f(\mathbf u^{(i)})^T\Delta\mathbf u$, we obtain
\[
\begin{aligned}
	&f(\mathbf u)-f(\mathbf u^{(i)})
	-\nabla f(\mathbf u^{(i)})^T\Delta\mathbf u\\
	&=
	\int_0^1(1-\tau)g''(\tau)\,d\tau\\
	&\geq
	-\chi_f^{(i)}\|\Delta\mathbf u\|_2^2
	\int_0^1(1-\tau)\,d\tau\\
	&=
	-\frac{\chi_f^{(i)}}{2}
	\|\Delta\mathbf u\|_2^2,
\end{aligned}
\]
where the inequality follows from $1-\tau\geq0$ on $[0,1]$. Rearranging establishes \eqref{48}.

The resulting quadratic lower bound $\underline f^{(i)}(\mathbf u)$ is concave because its Hessian is $-\chi_f^{(i)}\mathbf I_2\preceq\mathbf0$. At $\mathbf u=\mathbf u^{(i)}$, the displacement $\Delta\mathbf u$ is zero, giving
\[
\underline f^{(i)}(\mathbf u^{(i)})
=f(\mathbf u^{(i)}).
\]
Moreover,
\[
\nabla\underline f^{(i)}(\mathbf u)
=
\nabla f(\mathbf u^{(i)})
-\chi_f^{(i)}\Delta\mathbf u,
\]
and hence
$\nabla\underline f^{(i)}(\mathbf u^{(i)})
=\nabla f(\mathbf u^{(i)})$.
This completes the proof.

\bibliographystyle{IEEEtran}
\bibliography{Ref}
\vskip -2\baselineskip plus -2fil

\end{document}